\documentclass[journal=jpccck,manuscript=article]{achemso}

\usepackage[version=3]{mhchem} 
\usepackage[T1]{fontenc}       
\usepackage{amssymb}
\usepackage{tabularx,ragged2e}

\SectionNumbersOn

\usepackage{color, hyperref, multirow}

\author{Huan Tran}
\email{huan.tran@matmerize.com}
\affiliation{Matmerize Inc., 75 5th St NW, Atlanta, GA 30332, USA}
\alsoaffiliation{School of Materials Science and Engineering, Georgia Institute of Technology, 771 Ferst Dr. NW, Atlanta, GA 30332, USA}
\author{Akhlak Mahmood}
\affiliation{Matmerize Inc., 75 5th St NW, Atlanta, GA 30332, USA}
\author{Harshal Chaudhari}
\affiliation{Shell India Markets Pvt. Ltd., Bengaluru 560048, Karnataka, India}
\author{Kuldeep Mamtani}
\affiliation{Shell India Markets Pvt. Ltd., Bengaluru 560048, Karnataka, India}
\author{Chiho Kim}
\affiliation{Matmerize Inc., 75 5th St NW, Atlanta, GA 30332, USA}
\alsoaffiliation{School of Materials Science and Engineering, Georgia Institute of Technology, 771 Ferst Dr. NW, Atlanta, GA 30332, USA}
\author{Rampi Ramprasad}
\affiliation{Matmerize Inc., 75 5th St NW, Atlanta, GA 30332, USA}
\alsoaffiliation{School of Materials Science and Engineering, Georgia Institute of Technology, 771 Ferst Dr. NW, Atlanta, GA 30332, USA}
\author{Anand Krishnamoorthy}
\email{A.NarayananKrishnam2@shell.com}
\affiliation{Shell India Markets Pvt. Ltd., Bengaluru 560048, Karnataka, India}
\author{Abhirup Patra}
\email{Abhirup.Patra@shell.com }
\affiliation{Shell International Exploration \& Production Inc., 200 N. Dairy Ashford Rd., Houston, TX 77079, USA}

\title {Accelerated design of proton exchange membranes for green hydrogen production with artificial intelligence}

\begin{document}



\begin{abstract}
Water electrolysis is an eco-friendly method for hydrogen production that has reached significant levels of technological maturity. Among commercialized water-electrolysis technologies, proton-exchange membrane electrolyzers offer high current density, fast dynamic response, and compact system design, among other advantages. On the other hand, managing their high capital cost and the ``forever-chemistry'' nature of Nafion, a perfluorinated proton-exchange membrane widely used in such devices, remains a major challenge. Searches for fluorine-free replacements for Nafion, pursued largely through physical experimentation, have been active for decades with limited success. In this work, we develop and demonstrate an AI-based strategy for designing proton-exchange membranes for electrolyzers. Two key components of this strategy are an implementation of the virtual forward-synthesis approach and a set of machine-learning predictive models for essential application-inspired membrane properties; the former generates a vast space of millions of synthesizable polymers, which are then evaluated and screened by the latter. The strategy is validated against experimental data for known membranes and then applied to design over 1,700 synthesizable candidates. This article concludes with a forward-looking vision in which the strategy could be elevated into an interactive and iterative scheme that are based on large language models to facilitate materials design in multiple ways.

\end{abstract}
\newpage
\section{Introduction}
\begin{figure}[t]
\centering
\includegraphics[width=1.0\linewidth]{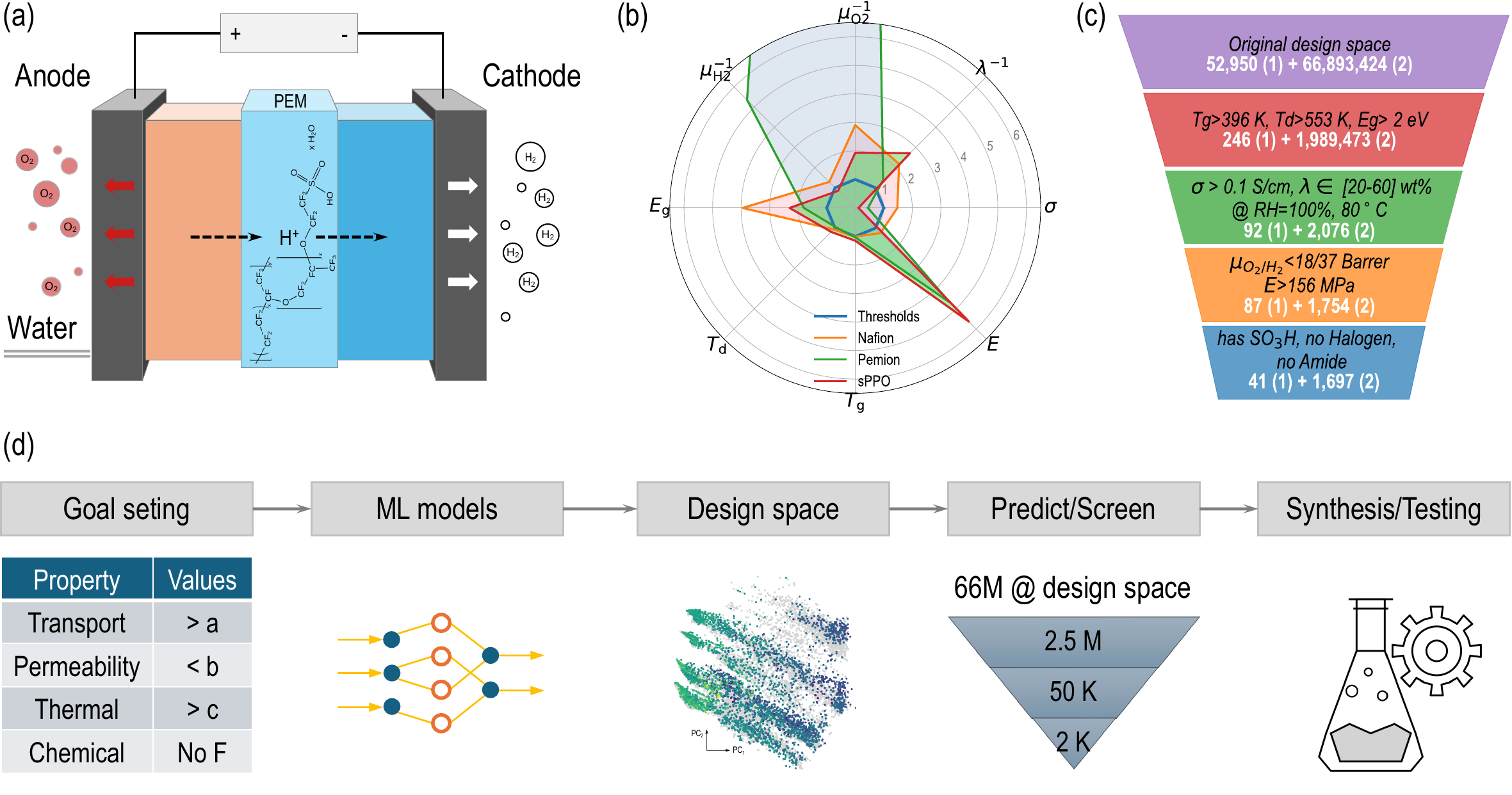}
	\caption{(a) A (schematic) cross-sectional view of a PEM electrolyzer in which a direct-current power source is used to split provided water molecules into hydrogen and oxygen gases in cathode and anode, (b) the chemical structure and a visualization of Nafion membrane, (c) the performances of Nafion, Pemion, and sPPO, given with respect to the application-inspired thresholds, and (d) the employed design strategy for PEMs. In (c), the design criteria are represented by a threshold polygon, outside of which good candidates for PEMs are expected to be.}\label{fig:cell}
\end{figure}

Hydrogen is an ideal ``fuel'' that can be efficiently converted to multiple forms of end-use energy, e.g., electricity (using fuel cells) and mechanical energy (using hydrogen internal combustion engines).\cite{zuttel2010hydrogen, abdalla2018hydrogen, akal2020review} At the industrial scale, hydrogen is primarily produced from natural gases using steam reforming, gasification, and pyrolysis, during which a great volume of carbon dioxide is also generated.\cite{mazloomi2012hydrogen, kumar2022overview} \textit{Electrolysis}, the process in which water molecules are decomposed into hydrogen (H$_2$) and oxygen (O$_2$) gases using electrical energy, is a \textit{green} method with essentially no carbon footprint.\cite{mansilha2026comprehensive, chatenet2022water, bodard2024green, wang2025proton, nagao2024proton, kumar2022overview} In an electrolytic cell, schematically shown in Fig. \ref{fig:cell} (a), a direct current runs between the anode and cathode, driving two electrochemical reactions, i.e., oxygen evolution reaction (OER) and hydrogen evolution reaction (HER), that create H$_2$ and O$_2$ from water. An electrolyte layer sandwiched between the electrodes enables the transport of designated ions (e.g., H$^+$ or OH$^-$) while preventing the crossover of electrons, H$_2$, and O$_2$. Categorized by the cell architecture and electrolyte, major types electrolyzers are alkaline water electrolyzer, proton-exchange membrane (PEM) water electrolyzer, anion exchange membrane water electrolyzer, and solid oxide electrolysis cells.\cite{chatenet2022water, kumar2022overview} Technically, water electrolysis is scalable and insusceptible to the intermittency and variability of renewable electricity, typically generated from wind, solar, hydro, geothermal, and tidal sources. Toward a green future, this hydrogen production method is attractive for both commercial and policy reasons.\cite{wang2025proton}

PEM electrolyzers are a commercialized technology, offering high current density, fast dynamic response, high pressure operation, compact system design, and long lifetime.\cite{kumar2022overview} Compared to alkaline water electrolyzers, another mature technology,\cite{krishnan2023present, hubert2024hydrogen} PEM electrolyzers are expensive because of their precious-metal electrocatalysts and proton-exchange membranes, typically Nafion [see Fig. \ref{fig:cell}(b)], known for its remarkable proton conductivity and durability under harsh working conditions. Beyond the capital cost considerations, recycling this perfluorinated sulphonic acid copolymer is economically and environmentally highly unfavorable due to its superior stability, which arises from the exceptionally strong carbon-fluorine bonds dominating the membrane.\cite{mansilha2026comprehensive, chatenet2022water, bodard2024green, wang2025proton, nagao2024proton, kumar2022overview} 

These long-standing issues motivate numerous active searches for sustainable (e.g., halogen-free), degradable, and more cost-effective PEMs during the last decades with limited successes.\cite{chatenet2022water, miyake2017design,souzy2005proton, wang2020fundamentals,kraytsberg2014review, tran2024design, Huan:fuel_cell, schertzer2025ai} The primary challenge is that the polymer space is prohibitively large for traditional methods, typically relying on laborious and time-consuming physical experimentations and computer simulations. Artificial intelligence (AI)-based methods have also been developed and introduced in the field.\cite{tran2024design, Huan:fuel_cell, schertzer2025ai, garcia2026comprehensive, chen2024machine, xu2024degradation, polo2025modeling, zhang2024data, ozdemir2024performance, zerrougui2025physics, makki2025machine, yu2026multiphysics} Recent AI efforts were devoted to various important PEM material properties and/or device performances,\cite{yu2026multiphysics} including proton conductivity and water uptake\cite{tran2024design, Huan:fuel_cell, schertzer2025ai}, mechanical, thermal, and permeability properties,\cite{Huan:fuel_cell} degradation\cite{chen2024machine, xu2024degradation, polo2025modeling}, voltage,\cite{zhang2024data, polo2025modeling} current density and hydrogen flow rate,\cite{ozdemir2024performance} operating temperature,\cite{zerrougui2025physics} and cathodic H$_2$ recovery.\cite{makki2025machine} Such promising AI approaches, while still limited in scale and scope for various reasons, could potentially accelerate the optimization of PEM materials and devices in the future. At the level of material discoveries, some halogen-free membranes, including Pemion\cite{nguyen2021hydrocarbon, nguyen2022fully, mirfarsi2024high} and sulfonated poly(2,6-dimethyl-1,4-phenylene oxide) (sPPO),\cite{ashcraft2010structure, afsar2019sppo, khan2022speek, yang2006sulfonated, xu2008poly, lee2016sppo, nagendra2023poly} have been identified, studied, and even commercialized.\cite{permionionomr, innochemtech} Compared to Nafion, their proton conductivity, stability, and/or durability under working conditions, as shown in Fig. \ref{fig:cell} (b), remain behind.\cite{nguyen2021hydrocarbon, nguyen2022fully, mirfarsi2024high}

This paper describes a large-scale, closed-loop AI-driven materials-design effort to develop halogen-free PEMs for water electrolyzers. The employed strategy, visualized in Fig. \ref{fig:cell} (d),\cite{tran2024design, Huan:fuel_cell} involves creating a set of 11 application-inspired property requirements for PEMs, developing the machine-learning (ML) models required to predict the properties,\cite{Chiho:PG, doan2020machine} generating a massive design space of millions of polymers using virtual forward synthesis (VFS) approach,\cite{kim2023open, ohno2023smipoly, gurnani2024ai, kern2025informatics} screening the space using the developed ML models for candidates, and testing them experimentally. After Nafion, Pemion, sPPO, and other sulfonate copolymers were ``rediscovered'' and validated, over 1,700 synthersizable polymers were identified from the screening, detailed in Fig. \ref{fig:cell} (c). For some of them, synthesis and testing are on-going and results are anticipated. The described AI-based polymer design strategy could be leveraged to develop an interactive, iterative large language model based scheme for future polymer research and development.

\section{Methods}
The critical steps of the employed AI-driven strategy, described below, are closely interconnected and driven by the application-specific requirements of the target material. The criteria must be derived from material properties for which reliable data are available. The technical criteria and sizing can only be stringent if the candidate space is sufficiently large; otherwise, no candidates can be effectively down-selected. As more and/or new data become available in the future, each step and the overall strategy can be further refined accordingly.

\subsection{Design criteria}\label{sec:designcriteria}
Taking the respective specifications of Nafion as reference, the property requirements for water electrolyzer PEMs were crafted and summarized in Table \ref{table:criteria}. Among them, a \textit{high} proton conductivity $\sigma$ is critically important. Because proton transport requires the presence of water molecules and a network of inter-connected water channels in PEMs,\cite{li2005principles, jiao2011water, kusoglu2017new, kusoglu2012water, hickner2005chemical} the proton conductivity $\sigma$ correlates strongly with the water uptake $\lambda$ while both $\sigma$ and $\lambda$ are functions of the temperature $T$ and the relative humidity RH of the working environment. For Nafion, high $\sigma$ can only be reached when $\lambda$ and RH are high, i.e., $\lambda \simeq 100$ wt\% and ${\rm RH} \simeq 100$\%, an undesirable requirement for practical operations.\cite{jiao2011water,kusoglu2017new,zhu2022recent} Therefore, a {\it moderate} level of water uptake $\lambda$, i.e., about 50 wt\% or below is targeted. Then, as the water channels must withstand internal pressure to remain connected during the membrane's lifetime, which could extend to tens of thousands of operating hours at $\simeq 80^\circ$C, PEMs should be mechanically strong and thermally robust. These requirements are translated into some thresholds that must be exceeded by the Young’s modulus $E$, the glass transition temperature $T_{\rm g}$, and the thermal decomposition temperature $T_{\rm d}$ (see Table \ref{table:criteria} for details).\cite{Huan:fuel_cell}

\begin{table*}[t]
\caption{Required performances of an electrolyzer's proton exchange membrane. For Young's modulus $E$, glass transition temperature $T_{\rm g}$, thermal decomposition temperature $T_{\rm d}$, O$_2$ permeability $\mu_{\rm O_2}$, and H$_2$ permeability $\mu_{\rm H_2}$, the thresholds reflect the actual performances of Nafion.}\label{table:criteria}
\begin{tabularx}{\textwidth}{@{} 
	>{\RaggedRight} p{0.5cm} 
	>{\RaggedRight} p{5.0cm}  
	>{\centering} p{1.5cm} 
	>{\centering} p{4.0cm} 
	>{\centering} p{4.0cm} 
	>{\RaggedRight} p{0.1cm}
	}
\hline
\hline
	No.	&Key property & Unit & Desired value & Nafion value& \\
\hline
	1 &       Proton conductivity $\sigma$ & S/cm    & $>0.1$ @ $T=80^\circ$C, RH=100\% & $\simeq 0.1$ @ $T=80^\circ$C, RH=100\% \cite{feng2018characterization, zhang2018sulfonated, peng2017preparation, si2012synthesis, wang2012clustered, silva2004tangential}&\\
	2	&       Water uptake $\lambda$               & wt\%    & $<50$ wt\% & $>50$ wt\% \cite{wang2012clustered, zhang2018sulfonated}&\\
	3	&	Young's modulus $E$ & MPa & $> 156$ & $50-220$\cite{zhang2018sulfonated, roberti2010measurement, caire2016accelerated} &\\
	4	&	Glass transition temperature $T_{\rm g}$ & K & $>396$ & $396-398$\cite{jung2012role, meyer2017investigation} &\\
	5	&	Thermal decomposition temperature $T_{\rm d}$ &K & $> 553$ & $553$\cite{lage2004thermal, samms1996thermal} &\\
	6	&	O$_2$ permeability $\mu_{\rm O_2}$ & Barrer& $< 18$ & $1.1-34.3$\cite{zhang2018sulfonated, mukaddam2016gas, fan2014role, chiou1988gas} &\\
	7	&	H$_2$ permeability $\mu_{\rm H_2}$ & Barrer & $< 37$ & $9.3-65.0$\cite{zhang2018sulfonated, mukaddam2016gas, fan2014role, chiou1988gas} &\\
	8	&	Electronic band gap $E_{\rm g}$ &eV & $> 2.0$ & N/A &\\
	9	&	Having halogen species  & N/A & No & Yes &\\
	10	&	Having amide -C(=O)NH- & N/A & No & No &\\
	11	&	Having sulfonate -SO$_3^-$ & N/A & Yes & Yes &\\
\hline
\hline
\end{tabularx}
\end{table*}

While promoting proton and water transports, the PEM layer in an electrolyzer should block the crossover of O$_2$ and H$_2$ gases. Therefore, the oxygen and hydrogen permeabilities, i.e., $\mu_{\rm O_2}$ and $\mu_{\rm H_2}$, of a good PEM should be low. Likewise, a PEM must be electronically insulating, and this requirement could be translated into a large enough electronic ``band gap'' $E_{\rm g}$. A high value of $E_{\rm g}$ is also needed to secure the electrochemical stability of the PEM.\cite{Huan:fuel_cell} As a voltage of about 1.23 V (on top of $\simeq 2.0$ eV between the anode and cathode in working conditions) is theoretically needed to split water molecules into O$_2$ and H$_2$,\cite{kumar2022overview} a criterion of $E_{\rm g} \geq 2.0$ eV  was used in this work.  For the Young's modulus $E$, glass transition temperature $T_{\rm g}$, thermal decomposition temperature $T_{\rm d}$, oxygen permeability $\mu_{\rm O_2}$, and hydrogen permeability $\mu_{\rm H_2}$, the thresholds, given in Table \ref{table:criteria}, approximate the reported performances of Nafion.

Some chemistry considerations were also considered for the design. First, avoiding halogen species is a primary objective, thus, the candidates should not have them. Then, amide groups -C(=O)NH- tend to make strong hydrogen bonds with water molecules, trapping them, and limiting the proton mobility. They are also prone to hydrolysis and can be a site for radical-based degradation, leading to long-term degradation of the membranes, typically soaked in water. Therefore, amide groups should also be excluded. Finally, a vast majority of the proton conductivity $\sigma$ and water uptake $\lambda$ data reported in the literature and curated for this project contains sulfonate (-SO$_3^-$), the hydrophilic functional groups that are essential for capturing water molecules in the membranes (see Sec. \ref{sec:data}).\cite{lowry1980investigation,kusoglu2017new, souzy2005proton} Consequently, we consider only candidates that have -SO$_3^-$ to make our $\sigma$ and $\lambda$ predictions reliable. Some other key materials-level properties such as the degradation of PEM under working conditions\cite{chen2024machine, xu2024degradation, polo2025modeling} are not considered here due to the current lack of sufficient and reliable data, but may be the subject of future work.

\begin{table*}[t]
\setlength{\tabcolsep}{1pt}
\caption{Summary of the dataset curated and the ML models trained on these datasets. The error metrics of the models are the coefficient of determination $R^2$ and either ``root-mean-square error'', given by the same unit with the data, or ``order of magnitude'', given by ``order'' and indicated by ``\underline{o}''.}\label{table:model}
\begin{tabularx}{\textwidth}{@{}
        >{\RaggedRight} p{2.4cm}
        >{\centering} p{1.2cm}
        >{\centering} p{2.2cm}
        >{\centering} p{5.8cm}
        >{\centering} p{1.2cm}
        >{\centering} p{2.5cm}
	>{\RaggedRight} p{0.1cm}
        }
\hline
\hline
	Property & Unit & Data size & Data range& $R^2$ & Error&\\
\hline
	$\sigma$                  & S/cm    & 2,462 &$5\times10^{-6}$ - $3.9\times10^{-1}$&0.84& $0.12$\underline{o} &\\
        $\lambda$                        & wt\%    & 2,120 &0.03 - 2,500&0.96& $0.10$\underline{o} &\\
        $E$                           & MPa     & 915   &0.02 - 4,000&0.82& $0.15$\underline{o} &\\
        $T_{\rm g}$      & K       & 8,962 &80 - 873&0.99& $10.0$  &\\
        $T_{\rm d}$ & K       & 6,585 &291 - 1,173&0.96& $21.9$  &\\
	$\mu_{\rm O_2}$            & Barrer  & 1,021&$1.5\times10^{-8}$ - $1.9\times10^{4}$&0.95&$0.07$\underline{o}&\\
	$\mu_{\rm H_2}$            & Barrer  & 603&$1.9\times10^{-6}$ - $3.7\times10^4$&0.97&$0.06$\underline{o}&\\
	$E_{\rm g}$               & eV      & 3,879 &0.1-9.8&0.97&$0.25$   &\\
\hline
\hline
\end{tabularx}
\end{table*}

\subsection{Training data and ML models}\label{sec:data}
Previously curated data\cite{Chiho:PG, doan2020machine, Huan:fuel_cell} were substantial expanded, creating the training datasets summarized in Table \ref{table:model}. The datasets of proton conductivity $\sigma$ and water uptake $\lambda$, two strongly related properties, contain 2,462 and 2,120 data points, respectively. About 97.8\% (2,409 entries) of the $\sigma$ dataset and 48.7\% (1,034 entries) of the $\lambda$ dataset involve -SO$_3^-$ sulfonate group, highlighting the focus of the field in the last several decades. Likewise, two datasets of O$_2$ and H$_2$ permeabilities are parts a bigger dataset, containing 4,377 entries of the permeabilities of six gases, including O$_2$, H$_2$, N$_2$, He, CO$_2$, a CH$_4$. Except for the band gap $E_{\rm g}$ dataset that was prepared by using density functional density calculations,\cite{Huan:Data, kamal2021novel, kamal2020computable} the other datasets are \textit{experimental} in nature. Polymers in these datasets are represented by {\sc smiles}, which stands for {\it simplified molecular-input line-entry system},\cite{smiles} from which a set of hierarchical chemical fingerprints\cite{Pilania_SR, Huan:design, Arun:design, Chiho:PG, doan2020machine} will be computed for the development of the ML models. For proton conductivity $\sigma$ and water uptake $\lambda$, the model development needs some additional descriptors, including the temperature $T$, the relative humidity RH, and another important measurement condition, i.e., whether the samples are submerged in liquid or vapor water. 

We used Gaussian Process Regression,\cite{GPRBook,GPR95} a similarity-based algorithm, integrating a radial basis function (RBF) kernel and an additive white-noise component, to develop the ML predictive models.\cite{Chiho:PG, doan2020machine} By assuming the output is a realization of a Gaussian process, the prediction uncertainty provided by GPR is useful for signaling whether a specific domain is well represented in the training data. An internal cross-validation mechanism was integrated to balance training and validation performance, thereby minimizing potential overfitting. As $\sigma$ and $\lambda$ are strongly correlated, a {\it multi-task} ML model was trained to predict both of them simultaneously. The rationale behind this solution is that for properties that are explicitly or implicitly related, multi-task algorithms are expected to exploit the possible correlations among them and make the models more robust and accurate.\cite{PATRA2020109286, kuenneth2021polymer, tran2025polymer} Likewise, another multi-task model was trained on the dataset combining the permeabilities of O$_2$, H$_2$, N$_2$, He, CO$_2$, a CH$_4$ for predicting $\mu_{\rm O_2}$ and $\mu_{\rm H_2}$ as needed for this strategy. Finally, four other ML models were trained to predict Young's modulus $E$, glass transition temperature $T_{\rm g}$, thermal decomposition temperature $T_{\rm d}$, and band gap $E_{\rm g}$. 

\subsection{Design space}
The design space (of homopolymers and copolymers) was constructed from about 30,000 known polymers,\cite{Chiho:PG, doan2020machine} i.e., those that have been synthesized, studied, and reported, and about 66 millions synthersizable polymers generated using RxnChainer,\cite{akhlak:rxnchainer} an implementation of the virtual forward synthesis (VFS) approach.\cite{kim2023open, ohno2023smipoly, gurnani2024ai, kern2025informatics} In a nutshell, VFS starts with millions of commercially available monomers, which can be found from multiple sources, including TSCA inventory,\cite{TSCA} ZINC-22,\cite{tingle2023zinc} ChemBL,\cite{mayr2018large} and eMolecules\cite{eMolecules}. These reactants are then propagated through numerous reaction chains, each of them is constructed from hundreds of rule-based polymerization templates, such as step growth, chain-growth addition, ring-opening, and metathesis, to generate the product polymers. In other words, VFS mimics the real polymerization reactions to create highly synthesizable polymers. Given the vast number of available monomers\cite{TSCA, tingle2023zinc, mayr2018large, eMolecules} and polymerization templates,\cite{gurnani2024ai, kern2025informatics} the number of polymers that can be generated using this approach is essentially unlimited.\cite{kim2023open, ohno2023smipoly, gurnani2024ai, kern2025informatics} Technical details of VFS and RxnChainer can be found in Ref. \citenum{akhlak:rxnchainer}.

The considered design space contains nearly 66 millions syntherizable and/or synthesized polymers, categorized into 2 subsets. Subset 1 has 52,950 polymers, including 26,958 polymer that have been synthesized, studied, and reported in the literature,\cite{Chiho:PG, doan2020machine} and 25,992 copolymers generated from known homopolymers. Subset 2 is much bigger, encompassing 66,893,424 homopolymers and copolymers generated from more than 7 millions commercial monomers obtained from TSCA inventory.\cite{TSCA} For practical reasons, four typical polymer families, namely polyimides, polyesters, polyureas, and polyurethanes, were considered for subset 2.

Results reported herein, including model training, design space creation, and the subsequent candidate screening (based on the established design criteria), were performed using PolymRize\texttrademark, a cloud-based polymer informatics software\cite{polymrize_url}. 

\section{Results}
\subsection{Machine-learning models}\label{sec:model}
The training process of the ML models is summarized in Table \ref{table:model}. Each trained ML model is characterized by a coefficient of determination $R^2$ and an error measure, which, depending on the training data distribution, can either be the root-mean-square error (RMSE) of the {\it order of magnitude error} (OME). For $T_{\rm g}$, $T_{\rm d}$, and $E_{\rm g}$, the data range is not too large, i.e., the upper bound is about 10 times of the lower bound or less, RMSE was used. On the other hand, as the data of $\sigma$, $\lambda$, $E$, $\mu_{\rm O_2}$, and $\mu_{\rm H_2}$ are highly scattered, i.e., spanning for 4-6 order of magnitudes, OME is a more suitable error metric. Formally, OME is defined as $\langle{\rm abs}(\log_{10}(p_{\rm pred}/p_{\rm ref}))\rangle$, where $p_{\rm pred}$ and $p_{\rm ref}$ are the predicted and reference values and $\langle\cdots\rangle$ stand for the average over the predictions. With this definition, OME provides the anticipated error of the predictions, given in ``orders of magnitude'' and denoted by \underline{o}. Both RMSE and OME should be examined in comparison with the range of the training data.

\begin{figure}[t]
\centering
\includegraphics[width=1.0\linewidth]{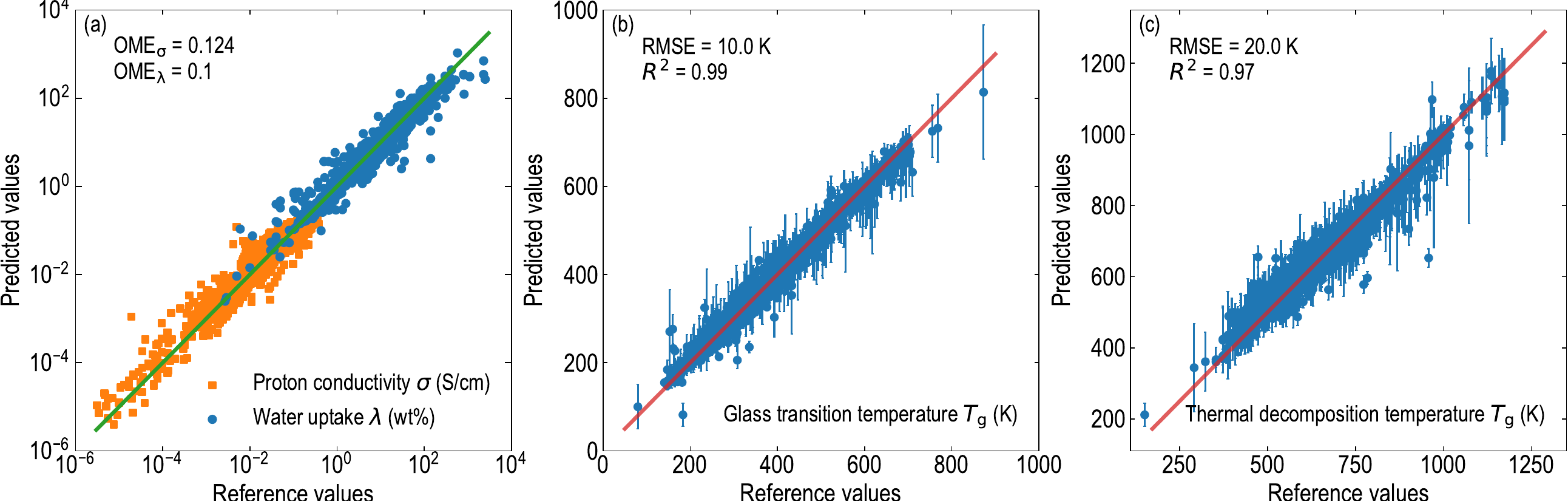}
\caption{ML models for (a) proton conductivity and water uptake, (b) glass transition temperature, and (c) thermal decomposition temperature.}\label{fig:model}
\end{figure}

The coefficient of determination $R^2$ of all the models is high. Except for $\sigma$ and $E$ models whose $R^2$ is 0.84 and 0.82, respectively, $R^2>0.95$ for other models. Two error measures, namely RMSE and OME, are small. In particular, RMSE obtained for $T_{\rm g}$, $T_{\rm d}$, and $E_{\rm g}$ models is about 3-7\% of the training data range while OME is about $0.1$\underline{o} for all the ML models developed for $\sigma$, $\lambda$, $E$, $\mu_{\rm O_2}$, and $\mu_{\rm H_2}$. This error measure is roughly 3-5\% of the training data range, which is about 4-6\underline{o}. Fig. \ref{fig:model} provides a visualization of the models trained for predicting $\lambda$, $\sigma$, $T_{\rm g}$, and $T_{\rm d}$. 

\subsection{Discovered PEM candidates}\label{sec:polymerspace}

As visualized in Figs. \ref{fig:cell} (c), a set of 1,738 candidates for electrolyzer PEMs were down-selected from the original design space, containing 41 polymers in subset 1 and 1,697 polymers (136 polyesters and 1,569 polyimides) in subset 2. While our experimental tests for sPPO were completed, synthesis and testing works for some other candidates are ongoing, and results may be anticipated.

\subsubsection{Known membranes: experimental validations}\label{sec:known_polymers}
\begin{figure}[t]
\centering
\includegraphics[width=1.0\linewidth]{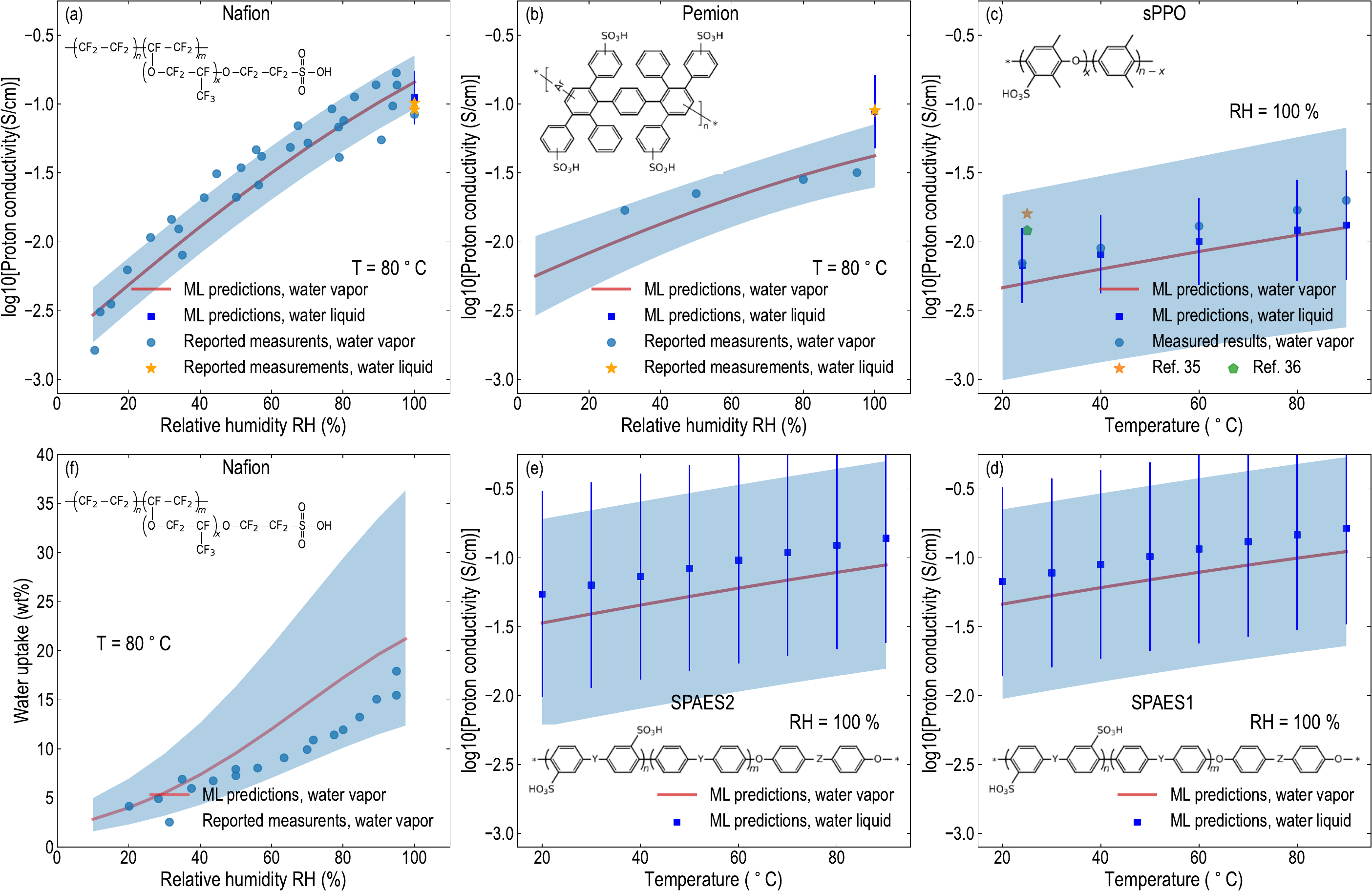}
\caption{Chemical structure, predicted proton conductivity (a) and water uptake (b) of Nafion, given in comparison with experimental data, taken from Refs. \citenum{feng2018characterization, zhang2018sulfonated, peng2017preparation, si2012synthesis, wang2012clustered, silva2004tangential} and Refs. \citenum{wang2012clustered, zhang2018sulfonated}, respectively. In (b), (c), (d), and (e), the predicted proton conductivity of Pemion, sPPO, SPAES1, and SPAES2 are given. Measured proton conductivity of Pemion was taken from Refs. \citenum{nguyen2021hydrocarbon, permionionomr}. For sPPO, $x=0.2$ and data were measured in this work. In SPAES1, Y is -S-, Z is -SO$_2$-, and $n/(m+n)=0.3$, while for SPAES2, Y is -S-, Z is -C(=O)-, and $n/(m+n)=0.3$. Shaded areas in (b)-(f) represent the uncertainty of the predictions for ``water vapor'' (given in solid lines).}\label{fig:sppo}
\end{figure}

Among the polymers identified from subset 1, Nafion, Pemion, and sPPO are examined in Figs. \ref{fig:sppo} (a), (b), and (c) with respect to available measured data. In Figs. \ref{fig:sppo} (a) and (b), the predicted proton conductivity captures very well the order and the trend of the experimental data previously reported for Nafion\cite{feng2018characterization, zhang2018sulfonated, peng2017preparation, si2012synthesis, wang2012clustered, silva2004tangential} and Pemion\cite{nguyen2021hydrocarbon, permionionomr} in both water vapor and water liquid measurement conditions, showing an excellent overall accuracy. For sPPO, we performed necessary measurements on the commercial membranes (InnoSep-C) acquired from Innochemtech Inc.\cite{innochemtech}. Using a BT-110 conductivity clamp, our electrochemical impedance spectroscopy measurements were performed to determine the high-frequency resistance $R$ at temperature ranging from 24 to 90$^\circ$C. Then, the in-plane proton conductivity $\sigma$ was calculated using  $\sigma = L/Rwt$ where $L$ is the distance between the reference electrodes while $w$ and $t$ are the width and thickness of the membrane. Our measured proton conductivity $\sigma$, together with data reported in Refs. \citenum{yang2006sulfonated, xu2008poly}, are presented Fig. \ref{fig:sppo} (c), providing another compelling confirmation of our ML predictions.

Ref. \citenum{McGrath2001ionconducting} claims several series of sulfonated aromatic poly(ether sulfone) colymers (SPAES), for one of them the generic chemical structure is shown in Figs. \ref{fig:sppo} (d) and (e). In this series, Y may come from a group consisting of -S-, S(O), -S(O)$_2$-, -C(O)-, -P(O)(C$_6$H$_5$)- and their combinations, Z may come from a group consisting of -C(CH$_3$)$_2$-, -C(CF$_3$)$_2$-, -C(CF$_3$)(C$_6$H$_5$)-, -C(O)-, -S(O)$_2$-, or -P(O)(C$_6$H$_5$)-, while $n/(m+n)$ may range from 0.001 to 1.\cite{McGrath2001ionconducting} By screening over all the non-halogen possibilities, we identified 2 candidates, denoted by SPAES1 and SPAES2, that meet our design criteria. SPAES1 includes -S- block for Y, -SO$_2$- block for Z, and $n/(m+n)=0.3$, while for SPAES2, Y is -S-, Z is -C(=O)-, and $n/(m+n)=0.3$.

\begin{table}[t]
	{\footnotesize
\setlength{\tabcolsep}{0.5pt}
	\caption{Summary of the agreements between predicted and measured data of Nafion, Pemion, and sPPO. References are given for the experimental data collected from the literature.}\label{table:sPPO}
    \centering
    \begin{tabular}{p{2.0cm} p{2.35cm} p{1.8cm} p{0.01cm} p{1.95cm} p{1.70cm} p{0.01cm} p{2.05cm} p{1.7cm} p{0.01cm} p{1.95cm} p{0.01cm} p{1.95cm}}
    \hline
    \hline
	    \multirow{2}{*}{Property} & \multicolumn{2}{c}{Nafion} && \multicolumn{2}{c}{Pemion} && \multicolumn{2}{c}{sPPO} && SPAES1 && SPAES2\\
	    \cline{2-3}\cline{5-6}\cline{8-9} \cline{11-11} \cline{13-13}
	     & Measured & Predicted && Measured & Predicted && Measured & Predicted && Predicted && Predicted\\
            \hline
	    $E$ (MPa) &50-220\cite{zhang2018sulfonated, roberti2010measurement, caire2016accelerated}& $195\pm 0.4$\underline{o}&&405-930\cite{mirfarsi2025mechanical}&$729\pm 0.6$\underline{o}&& N/A & $880\pm0.7$\underline{o} && $1,060\pm0.8$\underline{o} && $1,104\pm0.8$\underline{o}\\
	    $T_{\rm g}$ (K) &396-398\cite{jung2012role, meyer2017investigation}&$369\pm 32$&&$>393$\cite{mirfarsi2023thermo}&$408\pm33$&&$463\pm17$\cite{petreanu2012thermal,nagendra2023poly} & $455 \pm 13$  && $516\pm46$ && $517\pm45$\\
	    $T_{\rm d}$ (K) &553\cite{lage2004thermal, samms1996thermal}&$589\pm 18$&&$>533$\cite{pemion_td}&$529\pm10$&& 682\cite{petreanu2012thermal} & $658 \pm 18$ && $622\pm 48$ && $624\pm 47$ \\
	    $\mu_{\rm O_2}$ (Barrer) &1.1-34.3\cite{zhang2018sulfonated, mukaddam2016gas, fan2014role, chiou1988gas}&$6.2\pm 0.5$\underline{o}&&$\lesssim0.1\mu_{\rm O_2}^{\rm Naf}$\cite{garcia2023proton}&$0.3\pm 1.5$\underline{o}&& 6.4\cite{kruczek2001gas} & $9.3\pm 0.6$\underline{o} && $0.1\pm1.7$\underline{o} && $0.1\pm1.7$\underline{o} \\
	    $\mu_{\rm H_2}$ (Barrer) &9.3-65.0\cite{zhang2018sulfonated, mukaddam2016gas, fan2014role, chiou1988gas}&$28.7\pm 0.4$\underline{o}&&$\lesssim0.1\mu_{\rm H_2}^{\rm Naf}$\cite{garcia2023proton}&$6.9\pm0.9$\underline{o}&& 30.9\cite{kruczek2001gas} & $43.6\pm0.3$\underline{o} && $4.0\pm1.0$\underline{o} && $4.1\pm1.0$\underline{o} \\
	    $E_{\rm g}$ (eV)&N/A&$7.9\pm 0.2$&&N/A&$3.6\pm0.5$&& N/A & $4.6 \pm 0.4$ && $3.3 \pm 0.3$ && $3.3 \pm 0.3$ \\
         \hline
         \hline
    \end{tabular}}
\end{table}

The predicted water uptake of Nafion is shown in Figs. \ref{fig:sppo} (f). Compared to the measured data, obtained from Refs. \citenum{wang2012clustered, zhang2018sulfonated}, the predictions capture well both the order of magnitude and the trend of this important property, given as a function of the relative humidity. Nevertheless, the predictions shown in Figs. \ref{fig:sppo} (c)-(f) feature relatively high uncertainty, implying that the proton conductivity and water uptake models have not been exposed to enough volume of related data, and their training data should further be augmented. All the data predicted and measured for other properties are summarized in Table \ref{table:sPPO}.

Table \ref{table:sPPO} summarizes the predictions for other required performances of Nafion, Pemion, sPPO, SPAES1, and SPAES2. While the good agreement between the predicted and measured data of Nafion resembles our recent report,\cite{Huan:fuel_cell} similar agreements can also be found for Pemion and sPPO. The predicted glass transition temperature $T_{\rm g} = 485 \pm 26$ K and thermal decomposition temperature $T_{\rm d} = 619 \pm 59$ K agree very well with experimental values, namely $T_{\rm g} = 483$ K\cite{nagendra2023poly} and $T_{\rm d} = 682$ K.\cite{petreanu2012thermal} The prediction of Young's modulus $E = 746\pm0.8$\underline{o} MPa has no apparent experimental counterpart, but there is a report\cite{shiino2020structural} on three sulfonated polyphenylene-based copolymers with similar rigid backbone with reasonable measured Young's modulus of $E \simeq 1,100$ MPa. Likewise, within an uncertainty of about $0.3-0.6$\underline{o} (see Table \ref{table:model}), the predicted $\mu_{\rm O_2}$ and $\mu_{\rm H_2}$ are reasonably good compared to the experimental values. The predicted band gap $E_{\rm g} = 4.6 \pm 0.4$ eV indicates that sPPO is a good insulating material that could be suitable for the harsh working conditions of an electrolyzer. 

\subsubsection{Uncharted territory: PEM candidates}
Beyond the known polymers identified and validated in Sec. \ref{sec:known_polymers}, the subset of 1,697 candidates identified from the essentially unexplored space of 66M generated polymers are expected to yield interesting and useful discoveries. While experimental synthesis and testing for some of them are ongoing, we analyzed their chemical structure for critical functional groups and chemical features. 

\begin{figure}[t]
\centering
\includegraphics[width=1.0\linewidth]{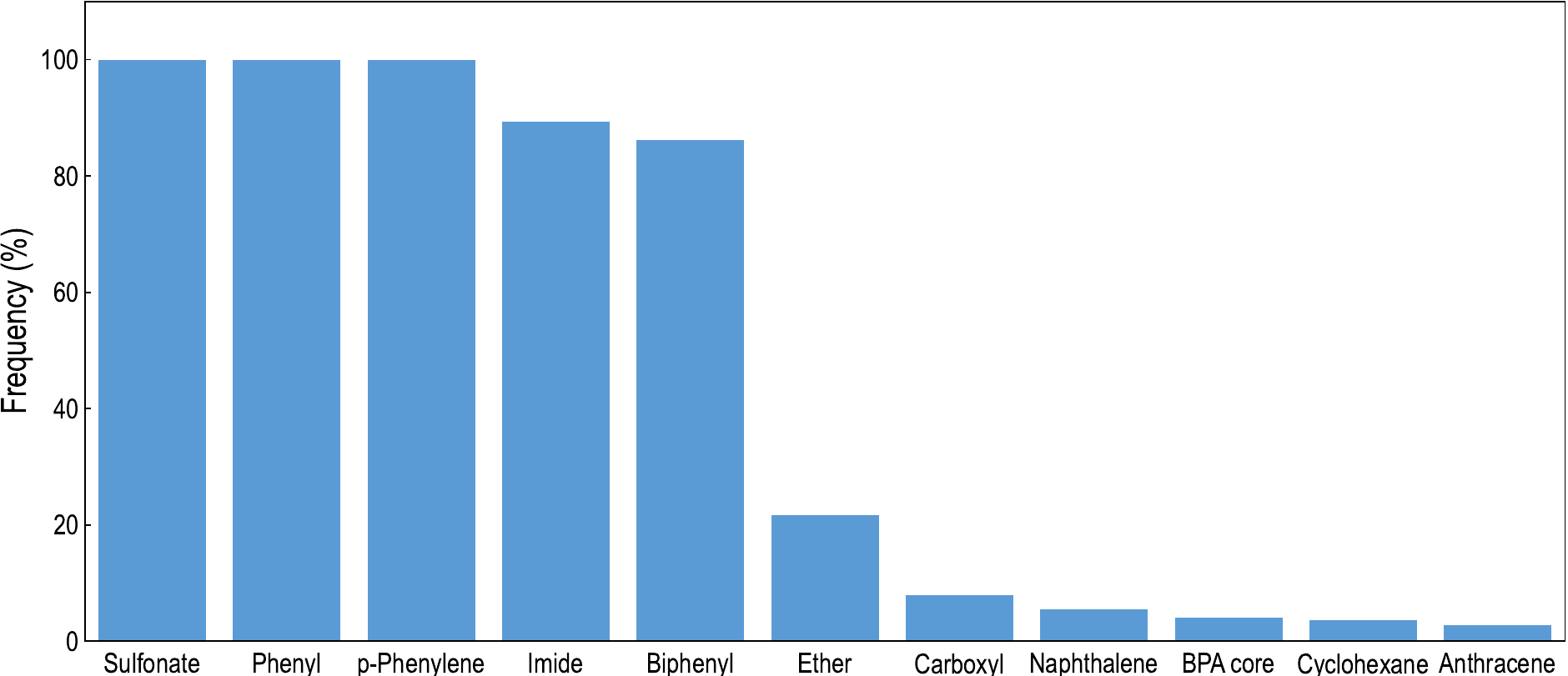}
\caption{Ten most-frequent functional groups/blocks found in 1,738 candidates identified from subset 2, containing 66M polymers generated using VFS.}\label{fig:stat}
\end{figure}

Starting from the design criteria, more than 50 functional groups (blocks) that may be relevant to any property were compiled. The chemical structure of the candidates, given in their SMILES, was then examined, returning 10 most-frequent blocks shown in Fig. \ref{fig:stat}. Because water molecules are critical for the dominant proton transport mechanisms in PEM, a vast majority of the reported literature data of proton conductivity $\sigma$ and water uptake $\lambda$ involve the hydrophilic pendant sulfonate functional groups -SO$_3^-$, needed for capturing and holding water molecules.\cite{lowry1980investigation,cable1995effects, korzeniewski2008responses,kusoglu2017new,souzy2005proton,liu2022comb} Therefore, this group is required as a screening criterion.\cite{Huan:fuel_cell} In the future, when data involving promising alternative functional groups such as sulfonimide (-SO$_2$NH-SO$_2$-)\cite{sutradhar2025synthesis,tsakanika2026single} become more abundant, the presence of sulfonate groups in the PEM candidates would be lower.

Phenyl (C$_6$H$_5$-) and 1,4-phenylene (-C$_6$H$_4$-) groups were found in every candidate, while biphenyl (-C$_6$H$_4$-C$_6$H$_4$-) occurs in 86\% of them. The very high rotational barrier of 1,4-phenylene and the bulky, rigid phenyl side group strongly suppress segmental mobility and raise the glass transition temperature $T_{\rm g}$. Likewise, C($sp^2$)-C($sp^2$) and aromatic C-H bonds are strong with high bond‐dissociation energies ($\sim110$ kcal/mol, versus $\sim100$ kcal/mol for aliphatic $sp^3$ bonds), thereby increasing the thermal decomposition temperature $T_{\rm d}$. Moreover, the quasi-planar geometry of phenyl and 1,4-phenylene also promotes tighter packing, limiting the crossover of gases such as O$_2$ and H$_2$, as required for a PEM. The imide group (-CO-NH-CO-), present in 89\% of the candidate set, possesses dipoles that improve cohesion and the thermal and mechanical properties of the membranes. Overall, sulfonate, phenyl, 1,4-phenylene, biphenyl, and imide are the most important groups of our candidate set, reasonably and collectively promoting the critical properties required for electrolyzer PEMs.

\section{Outlooks}
The AI-based application-driven design strategy, demonstrated herein and in recent works\cite{kim2026ai}, is inherently generic and opens several avenues for future polymer research and development. A particularly promising direction is the development of a large-language-model-based (AI) agent that integrates multi-objective optimization algorithms with the demonstrated VFS-ML workflow. Such an agent would enable interactive, application-driven polymer design while minimizing significant technical efforts currently required to train ML models, generate suitable polymer spaces, and identify optimal candidates. In this envisioned framework, users will interact with an AI agent through a natural language chat interface, specifying the desired performance metrics, constraints, and other requirements in conversational terms rather than through rigid parameter inputs. For example, a user may request ``a halogen-free membrane with proton conductivity similar to Nafion but with improved mechanical strength at elevated temperatures.'' The AI agent would then parse the requirements, translate them into quantitative design criteria, e.g., $\sigma > 0.1$ S/cm at 80$^\circ$C, $E > 200$ MPa, $T_{\rm g} > 450$ K, and leverage the VFS-ML framework to generate and evaluate the candidate materials. 

\begin{figure}[t]
\centering
\includegraphics[width=1.0\linewidth]{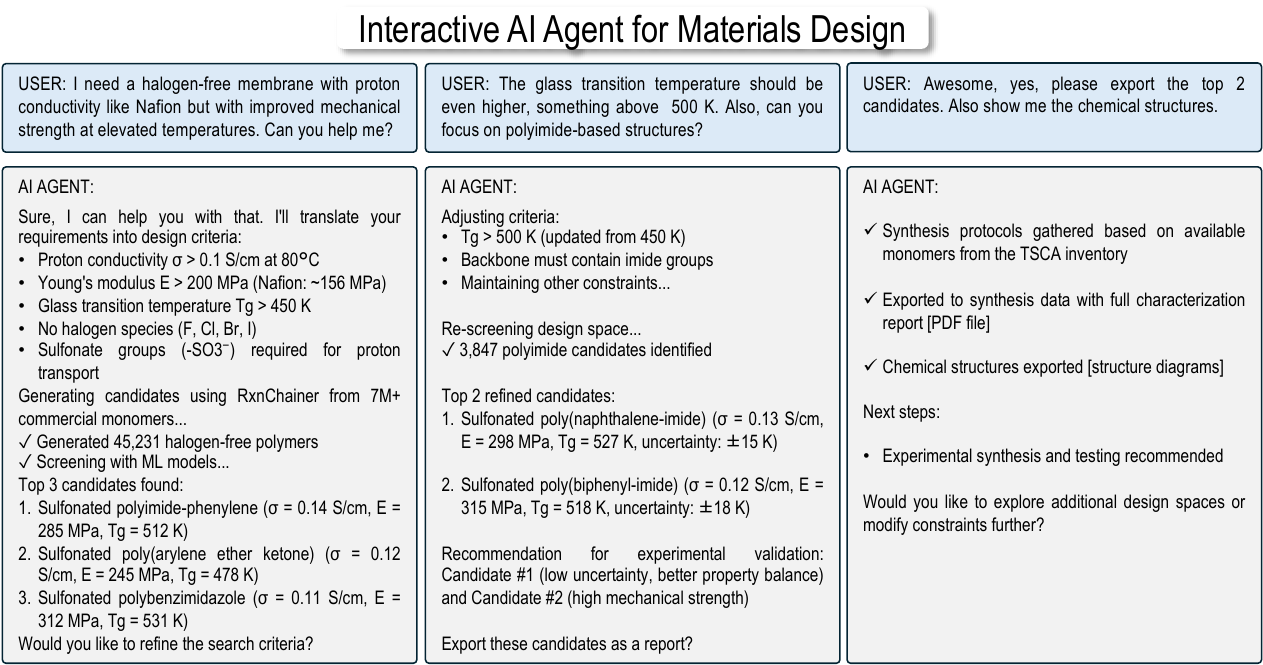}
	\caption{Schematic illustration of the envisioned interactive AI agent workflow for application-driven polymer design. User specify performance requirements through a natural language interface, the AI agent translates the requirements into quantitative design criteria, generates polymer candidates using RxnChainer, predicts their properties using trained ML models, and engages in iterative refinement loop based on user feedback. The agent provides explanations of structure-property relationships and presents ranked candidates with predicted property values, uncertainty estimates, and synthesizability scores.}\label{fig:aiagent}
\end{figure}

An example of such an interaction is illustrated in Fig. \ref{fig:aiagent}. Having an user request, the agent would utilize RxnChainer\cite{akhlak:rxnchainer} to generate polymer candidates from the vast space of commercially available monomers, applying the user-specified chemical constraints (e.g., excluding halogen species, requiring sulfonate groups). The candidates would then be screened using the trained ML models to predict their properties. Critically, the agent would not simply return a ranked list of candidates but also engage in an iterative refinement process based on user feedback. If the initial candidates do not meet expectations, the user could provide additional guidance, e.g., ``increase the glass transition temperature threshold'' or ``focus on polyimide-based structures'', and the agent would adjust the search accordingly. Throughout this process, the agent would provide explanations of the underlying structure-property relationships, highlighting which chemical moieties contribute to specific properties. For instance, it might explain that ``the presence of rigid aromatic rings in the backbone enhances $T_{\rm g}$ and $E$, while the sulfonate side groups enable high $\sigma$ through water uptake.''

Finally, the agent could also suggest experimental validation priorities based on the prediction uncertainties and potential impact. This interactive and iterative approach could transform the polymer design process from a one-time screening exercise into a dynamic collaboration between human expertise and AI capabilities. Such an intelligent framework could significantly accelerate the discovery of high-performance materials not only for PEMs but for a broad range of functional polymer applications, making advanced materials design accessible to researchers without extensive ML expertise. An experimental version of such an agentic capability (ASKPOLY) is available.\cite{askpoly} 

\section{Summary}
We have demonstrated an AI-based, application-driven design scheme to develop sustainable (halogen-free) proton-exchange membranes for PEM electrolyzers. Central to this strategy are a virtual forward-synthesis approach, used to generate a large design space of synthesizable polymers, and a set of ML predictive models developed to evaluate candidates within this space. The ML models were validated against experimental data curated and/or generated for Nafion, Pemion, and sPPO. The strategy was then used to discover 1,738 promising PEM candidates, some of which are undergoing experimental synthesis and testing, with further results anticipated. This strategy also suggests avenues for future polymer research and development. A particularly promising direction is the development of an AI agent that enables seamless interactions between users and the demonstrated AI-based strategy, fostering interactive and iterative human–machine collaboration for accelerated materials design.

\section*{Data Availability Statement}
The data used for this work was collected from public sources cited in the manuscript. Derivatives and data analysis results that support the discussions and conclusions of this work are available in the main text. 


\begin{acknowledgement}
This work was done within the GameChanger Initiative of Shell. The authors extend their sincere appreciation to Sreenivas Raghavendran for discussions and supports throughout the project. They gratefully acknowledge Innochemtech Co. Ltd., Republic of Korea, for generously providing the membrane samples free of charge, and especially Dr. Won-keun Son (CEO), Ji-won Kim, and Dr. S. Kumar for their support and coordination. The authors would also like to thank Sajanikumari Sadasivan and Cor van Kruijsdijk for their guidance on domain expertise.

\end{acknowledgement}

\vspace{6mm}
\noindent \textbf{Notes}
\noindent Competing interests: The authors declare no competing financial interests.




\begin{mcitethebibliography}{101}
\providecommand*\natexlab[1]{#1}
\providecommand*\mciteSetBstSublistMode[1]{}
\providecommand*\mciteSetBstMaxWidthForm[2]{}
\providecommand*\mciteBstWouldAddEndPuncttrue
  {\def\EndOfBibitem{\unskip.}}
\providecommand*\mciteBstWouldAddEndPunctfalse
  {\let\EndOfBibitem\relax}
\providecommand*\mciteSetBstMidEndSepPunct[3]{}
\providecommand*\mciteSetBstSublistLabelBeginEnd[3]{}
\providecommand*\EndOfBibitem{}
\mciteSetBstSublistMode{f}
\mciteSetBstMaxWidthForm{subitem}{(\alph{mcitesubitemcount})}
\mciteSetBstSublistLabelBeginEnd
  {\mcitemaxwidthsubitemform\space}
  {\relax}
  {\relax}

\bibitem[Z{\"u}ttel \latin{et~al.}(2010)Z{\"u}ttel, Remhof, Borgschulte, and
  Friedrichs]{zuttel2010hydrogen}
Z{\"u}ttel,~A.; Remhof,~A.; Borgschulte,~A.; Friedrichs,~O. Hydrogen: the
  future energy carrier. \emph{Philos. Trans. A or Phil. Trans. R. Soc. A}
  \textbf{2010}, \emph{368}, 3329--3342\relax
\mciteBstWouldAddEndPuncttrue
\mciteSetBstMidEndSepPunct{\mcitedefaultmidpunct}
{\mcitedefaultendpunct}{\mcitedefaultseppunct}\relax
\EndOfBibitem
\bibitem[Abdalla \latin{et~al.}(2018)Abdalla, Hossain, Nisfindy, Azad, Dawood,
  and Azad]{abdalla2018hydrogen}
Abdalla,~A.~M.; Hossain,~S.; Nisfindy,~O.~B.; Azad,~A.~T.; Dawood,~M.;
  Azad,~A.~K. Hydrogen production, storage, transportation and key challenges
  with applications: A review. \emph{Energy Convers. Manag.} \textbf{2018},
  \emph{165}, 602--627\relax
\mciteBstWouldAddEndPuncttrue
\mciteSetBstMidEndSepPunct{\mcitedefaultmidpunct}
{\mcitedefaultendpunct}{\mcitedefaultseppunct}\relax
\EndOfBibitem
\bibitem[Akal \latin{et~al.}(2020)Akal, {\"O}ztuna, and
  B{\"u}y{\"u}kak{\i}n]{akal2020review}
Akal,~D.; {\"O}ztuna,~S.; B{\"u}y{\"u}kak{\i}n,~M.~K. A review of hydrogen
  usage in internal combustion engines (gasoline-Lpg-diesel) from combustion
  performance aspect. \emph{Int. J. Hydrogen Energy} \textbf{2020}, \emph{45},
  35257--35268\relax
\mciteBstWouldAddEndPuncttrue
\mciteSetBstMidEndSepPunct{\mcitedefaultmidpunct}
{\mcitedefaultendpunct}{\mcitedefaultseppunct}\relax
\EndOfBibitem
\bibitem[Mazloomi and Gomes(2012)Mazloomi, and Gomes]{mazloomi2012hydrogen}
Mazloomi,~K.; Gomes,~C. Hydrogen as an energy carrier: Prospects and
  challenges. \emph{Renew. Sustain. Energy Rev.} \textbf{2012}, \emph{16},
  3024--3033\relax
\mciteBstWouldAddEndPuncttrue
\mciteSetBstMidEndSepPunct{\mcitedefaultmidpunct}
{\mcitedefaultendpunct}{\mcitedefaultseppunct}\relax
\EndOfBibitem
\bibitem[Kumar and Lim(2022)Kumar, and Lim]{kumar2022overview}
Kumar,~S.~S.; Lim,~H. An overview of water electrolysis technologies for green
  hydrogen production. \emph{Energy Rep.} \textbf{2022}, \emph{8},
  13793--13813\relax
\mciteBstWouldAddEndPuncttrue
\mciteSetBstMidEndSepPunct{\mcitedefaultmidpunct}
{\mcitedefaultendpunct}{\mcitedefaultseppunct}\relax
\EndOfBibitem
\bibitem[Mansilha \latin{et~al.}(2026)Mansilha, Barbosa-P{\'o}voa, Tarelho, and
  Fonseca]{mansilha2026comprehensive}
Mansilha,~C.; Barbosa-P{\'o}voa,~A.; Tarelho,~L.; Fonseca,~A. A comprehensive
  review of green hydrogen production technologies: current status, challenges,
  research trends and future directions. \emph{Renew. Sustain. Energy Rev.}
  \textbf{2026}, \emph{225}, 116119\relax
\mciteBstWouldAddEndPuncttrue
\mciteSetBstMidEndSepPunct{\mcitedefaultmidpunct}
{\mcitedefaultendpunct}{\mcitedefaultseppunct}\relax
\EndOfBibitem
\bibitem[Chatenet \latin{et~al.}(2022)Chatenet, Pollet, Dekel, Dionigi,
  Deseure, Millet, Braatz, Bazant, Eikerling, Staffell, \latin{et~al.}
  others]{chatenet2022water}
Chatenet,~M.; Pollet,~B.~G.; Dekel,~D.~R.; Dionigi,~F.; Deseure,~J.;
  Millet,~P.; Braatz,~R.~D.; Bazant,~M.~Z.; Eikerling,~M.; Staffell,~I.
  \latin{et~al.}  Water electrolysis: from textbook knowledge to the latest
  scientific strategies and industrial developments. \emph{Chem. Soc. Rev.}
  \textbf{2022}, \emph{51}, 4583--4762\relax
\mciteBstWouldAddEndPuncttrue
\mciteSetBstMidEndSepPunct{\mcitedefaultmidpunct}
{\mcitedefaultendpunct}{\mcitedefaultseppunct}\relax
\EndOfBibitem
\bibitem[Bodard \latin{et~al.}(2024)Bodard, Chen, ELJarray, and
  Zhang]{bodard2024green}
Bodard,~A.; Chen,~Z.; ELJarray,~O.; Zhang,~G. Green Hydrogen Production by
  Low-Temperature Membrane-Engineered Water Electrolyzers, and Regenerative
  Fuel Cells. \emph{Small Methods} \textbf{2024}, \emph{8}, 2400574\relax
\mciteBstWouldAddEndPuncttrue
\mciteSetBstMidEndSepPunct{\mcitedefaultmidpunct}
{\mcitedefaultendpunct}{\mcitedefaultseppunct}\relax
\EndOfBibitem
\bibitem[Wang \latin{et~al.}(2025)Wang, Stansberry, Mukundan, Chang, Kulkarni,
  Park, Plymill, Firas, Liu, Lang, \latin{et~al.} others]{wang2025proton}
Wang,~C.~R.; Stansberry,~J.~M.; Mukundan,~R.; Chang,~H.-M.~J.; Kulkarni,~D.;
  Park,~A.~M.; Plymill,~A.~B.; Firas,~N.~M.; Liu,~C.~P.; Lang,~J.~T.
  \latin{et~al.}  Proton exchange membrane (PEM) water electrolysis: cell-level
  considerations for gigawatt-scale deployment. \emph{Chem. Rev.}
  \textbf{2025}, \emph{125}, 1257--1302\relax
\mciteBstWouldAddEndPuncttrue
\mciteSetBstMidEndSepPunct{\mcitedefaultmidpunct}
{\mcitedefaultendpunct}{\mcitedefaultseppunct}\relax
\EndOfBibitem
\bibitem[Nagao(2024)]{nagao2024proton}
Nagao,~Y. Proton-conducting polymers: key to next-generation fuel cells,
  electrolyzers, batteries, actuators, and sensors. \emph{ChemElectroChem}
  \textbf{2024}, \emph{11}, e202300846\relax
\mciteBstWouldAddEndPuncttrue
\mciteSetBstMidEndSepPunct{\mcitedefaultmidpunct}
{\mcitedefaultendpunct}{\mcitedefaultseppunct}\relax
\EndOfBibitem
\bibitem[Krishnan \latin{et~al.}(2023)Krishnan, Koning, de~Groot, de~Groot,
  Mendoza, Junginger, and Kramer]{krishnan2023present}
Krishnan,~S.; Koning,~V.; de~Groot,~M.~T.; de~Groot,~A.; Mendoza,~P.~G.;
  Junginger,~M.; Kramer,~G.~J. Present and future cost of alkaline and PEM
  electrolyser stacks. \emph{Int. J. Hydrogen Energy} \textbf{2023}, \emph{48},
  32313--32330\relax
\mciteBstWouldAddEndPuncttrue
\mciteSetBstMidEndSepPunct{\mcitedefaultmidpunct}
{\mcitedefaultendpunct}{\mcitedefaultseppunct}\relax
\EndOfBibitem
\bibitem[Hubert \latin{et~al.}(2024)Hubert, Esposito, Peterson, Miller, and
  Stanford]{hubert2024hydrogen}
Hubert,~M.; Esposito,~A.; Peterson,~D.; Miller,~E.; Stanford,~J. \emph{Hydrogen
  Shot: Water Electrolysis Technology Assessment}; US Department of Energy,
  2024\relax
\mciteBstWouldAddEndPuncttrue
\mciteSetBstMidEndSepPunct{\mcitedefaultmidpunct}
{\mcitedefaultendpunct}{\mcitedefaultseppunct}\relax
\EndOfBibitem
\bibitem[Miyake \latin{et~al.}(2017)Miyake, Taki, Mochizuki, Shimizu, Akiyama,
  Uchida, and Miyatake]{miyake2017design}
Miyake,~J.; Taki,~R.; Mochizuki,~T.; Shimizu,~R.; Akiyama,~R.; Uchida,~M.;
  Miyatake,~K. Design of Flexible Polyphenylene Proton-Conducting Membrane for
  Next-Generation Fuel Cells. \emph{Sci. Adv.} \textbf{2017}, \emph{3},
  eaao0476\relax
\mciteBstWouldAddEndPuncttrue
\mciteSetBstMidEndSepPunct{\mcitedefaultmidpunct}
{\mcitedefaultendpunct}{\mcitedefaultseppunct}\relax
\EndOfBibitem
\bibitem[Souzy \latin{et~al.}(2005)Souzy, Ameduri, Boutevin, Capron, Marsacq,
  and Gebel]{souzy2005proton}
Souzy,~R.; Ameduri,~B.; Boutevin,~B.; Capron,~P.; Marsacq,~D.; Gebel,~G.
  Proton-Conducting Polymer Electrolyte Membranes Based on Fluoropolymers
  Incorporating Perfluorovinyl Ether Sulfonic Acids and Fluoroalkenes:
  Synthesis and Characterization. \emph{Fuel Cells} \textbf{2005}, \emph{5},
  383--397\relax
\mciteBstWouldAddEndPuncttrue
\mciteSetBstMidEndSepPunct{\mcitedefaultmidpunct}
{\mcitedefaultendpunct}{\mcitedefaultseppunct}\relax
\EndOfBibitem
\bibitem[Wang \latin{et~al.}(2020)Wang, Seo, Wang, Zamel, Jiao, and
  Adroher]{wang2020fundamentals}
Wang,~Y.; Seo,~B.; Wang,~B.; Zamel,~N.; Jiao,~K.; Adroher,~X.~C. Fundamentals,
  Materials, and Machine Learning of Polymer Electrolyte Membrane Fuel Cell
  Technology. \emph{Energy and AI} \textbf{2020}, \emph{1}, 100014\relax
\mciteBstWouldAddEndPuncttrue
\mciteSetBstMidEndSepPunct{\mcitedefaultmidpunct}
{\mcitedefaultendpunct}{\mcitedefaultseppunct}\relax
\EndOfBibitem
\bibitem[Kraytsberg and Ein-Eli(2014)Kraytsberg, and
  Ein-Eli]{kraytsberg2014review}
Kraytsberg,~A.; Ein-Eli,~Y. Review of Advanced Materials for Proton Exchange
  Membrane Fuel Cells. \emph{Energy \& Fuels} \textbf{2014}, \emph{28},
  7303--7330\relax
\mciteBstWouldAddEndPuncttrue
\mciteSetBstMidEndSepPunct{\mcitedefaultmidpunct}
{\mcitedefaultendpunct}{\mcitedefaultseppunct}\relax
\EndOfBibitem
\bibitem[Tran \latin{et~al.}(2024)Tran, Gurnani, Kim, Pilania, Kwon, Lively,
  and Ramprasad]{tran2024design}
Tran,~H.; Gurnani,~R.; Kim,~C.; Pilania,~G.; Kwon,~H.-K.; Lively,~R.~P.;
  Ramprasad,~R. Design of functional and sustainable polymers assisted by
  artificial intelligence. \emph{Nat. Rev. Mater.} \textbf{2024}, \emph{9},
  866--886\relax
\mciteBstWouldAddEndPuncttrue
\mciteSetBstMidEndSepPunct{\mcitedefaultmidpunct}
{\mcitedefaultendpunct}{\mcitedefaultseppunct}\relax
\EndOfBibitem
\bibitem[Tran \latin{et~al.}(2023)Tran, Shen, Shukla, Kwon, and
  Ramprasad]{Huan:fuel_cell}
Tran,~H.; Shen,~K.-H.; Shukla,~S.; Kwon,~H.-K.; Ramprasad,~R.
  Informatics-Driven Selection of Polymers for Fuel-Cell Applications. \emph{J.
  Phys. Chem. C} \textbf{2023}, \emph{127}, 977--986\relax
\mciteBstWouldAddEndPuncttrue
\mciteSetBstMidEndSepPunct{\mcitedefaultmidpunct}
{\mcitedefaultendpunct}{\mcitedefaultseppunct}\relax
\EndOfBibitem
\bibitem[Schertzer \latin{et~al.}(2025)Schertzer, Shukla, Sose, Rafiq, Al~Otmi,
  Sampath, Lively, and Ramprasad]{schertzer2025ai}
Schertzer,~W.; Shukla,~S.; Sose,~A.; Rafiq,~R.; Al~Otmi,~M.; Sampath,~J.;
  Lively,~R.~P.; Ramprasad,~R. AI-driven design of fluorine-free polymers for
  sustainable and high-performance anion exchange membranes. \emph{J. Mater.
  Infor.} \textbf{2025}, \emph{5}, N--A\relax
\mciteBstWouldAddEndPuncttrue
\mciteSetBstMidEndSepPunct{\mcitedefaultmidpunct}
{\mcitedefaultendpunct}{\mcitedefaultseppunct}\relax
\EndOfBibitem
\bibitem[García-Salaberri \latin{et~al.}(2026)García-Salaberri, Sanumi,
  Ares~de Parga-Regalado, Rozo, Blanco-Carmona, van~der Heijden, McKague,
  Mehrnia, Batool, Barzegari, Bordons, Iranzo, Ryzhakov, Gostick,
  Forner-Cuenca, Torralba, Jankovic, and Secanell]{garcia2026comprehensive}
García-Salaberri,~P.~A.; Sanumi,~O.; Ares~de Parga-Regalado,~A.~M.;
  Rozo,~D.~F.; Blanco-Carmona,~P.; van~der Heijden,~M.; McKague,~M.;
  Mehrnia,~M.; Batool,~M.; Barzegari,~M. \latin{et~al.}  A Comprehensive Review
  of Machine Learning in Proton-Exchange Membrane Fuel Cells: From Materials to
  Intelligent Systems. \emph{Energy \& Environmental Science} \textbf{2026},
  Submitted\relax
\mciteBstWouldAddEndPuncttrue
\mciteSetBstMidEndSepPunct{\mcitedefaultmidpunct}
{\mcitedefaultendpunct}{\mcitedefaultseppunct}\relax
\EndOfBibitem
\bibitem[Chen \latin{et~al.}(2024)Chen, Rex, Woelke, Eckert, Bensmann,
  Hanke-Rauschenbach, and Geyer]{chen2024machine}
Chen,~X.; Rex,~A.; Woelke,~J.; Eckert,~C.; Bensmann,~B.;
  Hanke-Rauschenbach,~R.; Geyer,~P. Machine learning in proton exchange
  membrane water electrolysis—A knowledge-integrated framework. \emph{Appl.
  Energy} \textbf{2024}, \emph{371}, 123550\relax
\mciteBstWouldAddEndPuncttrue
\mciteSetBstMidEndSepPunct{\mcitedefaultmidpunct}
{\mcitedefaultendpunct}{\mcitedefaultseppunct}\relax
\EndOfBibitem
\bibitem[Xu \latin{et~al.}(2024)Xu, Ma, Wu, Wang, Yang, Li, Zhu, and
  Liao]{xu2024degradation}
Xu,~B.; Ma,~W.; Wu,~W.; Wang,~Y.; Yang,~Y.; Li,~J.; Zhu,~X.; Liao,~Q.
  Degradation prediction of PEM water electrolyzer under constant and
  start-stop loads based on CNN-LSTM. \emph{Energy and AI} \textbf{2024},
  \emph{18}, 100420\relax
\mciteBstWouldAddEndPuncttrue
\mciteSetBstMidEndSepPunct{\mcitedefaultmidpunct}
{\mcitedefaultendpunct}{\mcitedefaultseppunct}\relax
\EndOfBibitem
\bibitem[Polo-Molina \latin{et~al.}(2025)Polo-Molina, Portela, Rozas, and
  Gonz{\'a}lez]{polo2025modeling}
Polo-Molina,~A.; Portela,~J.; Rozas,~L. A.~H.; Gonz{\'a}lez,~R.~C. Modeling
  membrane degradation in PEM electrolyzers with physics-informed neural
  networks. \emph{arXiv preprint arXiv:2507.02887} \textbf{2025}, \relax
\mciteBstWouldAddEndPunctfalse
\mciteSetBstMidEndSepPunct{\mcitedefaultmidpunct}
{}{\mcitedefaultseppunct}\relax
\EndOfBibitem
\bibitem[Zhang \latin{et~al.}(2024)Zhang, Tan, Yuan, Zhao, Shi, Liu, and
  Liu]{zhang2024data}
Zhang,~Y.; Tan,~A.; Yuan,~Z.; Zhao,~K.; Shi,~X.; Liu,~P.; Liu,~J. Data-driven
  optimization of high-dimensional variables in proton exchange membrane water
  electrolysis membrane electrode assembly assisted by machine learning.
  \emph{Ind. Eng. Chem. Res.} \textbf{2024}, \emph{63}, 1409--1421\relax
\mciteBstWouldAddEndPuncttrue
\mciteSetBstMidEndSepPunct{\mcitedefaultmidpunct}
{\mcitedefaultendpunct}{\mcitedefaultseppunct}\relax
\EndOfBibitem
\bibitem[Ozdemir and Pektezel(2024)Ozdemir, and
  Pektezel]{ozdemir2024performance}
Ozdemir,~S.~N.; Pektezel,~O. Performance prediction of experimental PEM
  electrolyzer using machine learning algorithms. \emph{Fuel} \textbf{2024},
  \emph{378}, 132853\relax
\mciteBstWouldAddEndPuncttrue
\mciteSetBstMidEndSepPunct{\mcitedefaultmidpunct}
{\mcitedefaultendpunct}{\mcitedefaultseppunct}\relax
\EndOfBibitem
\bibitem[Zerrougui \latin{et~al.}(2025)Zerrougui, Li, and
  Hissel]{zerrougui2025physics}
Zerrougui,~I.; Li,~Z.; Hissel,~D. Physics-Informed Neural Network for modeling
  and predicting temperature fluctuations in proton exchange membrane
  electrolysis. \emph{Energy and AI} \textbf{2025}, \emph{20}, 100474\relax
\mciteBstWouldAddEndPuncttrue
\mciteSetBstMidEndSepPunct{\mcitedefaultmidpunct}
{\mcitedefaultendpunct}{\mcitedefaultseppunct}\relax
\EndOfBibitem
\bibitem[Makki~Abadi and Rashidi(2025)Makki~Abadi, and
  Rashidi]{makki2025machine}
Makki~Abadi,~M.; Rashidi,~M.~M. Machine learning for the optimization and
  performance prediction of solid oxide electrolysis cells: a review.
  \emph{Processes} \textbf{2025}, \emph{13}, 875\relax
\mciteBstWouldAddEndPuncttrue
\mciteSetBstMidEndSepPunct{\mcitedefaultmidpunct}
{\mcitedefaultendpunct}{\mcitedefaultseppunct}\relax
\EndOfBibitem
\bibitem[Yu \latin{et~al.}(2026)Yu, Luo, Han, Mao, and Liu]{yu2026multiphysics}
Yu,~C.; Luo,~L.; Han,~Y.; Mao,~P.; Liu,~Y. Multiphysics and Multiscale Modeling
  of PEM Water Electrolyzers: From Transport Mechanisms to Performance
  Optimization. \emph{Energies} \textbf{2026}, \emph{19}, 2361\relax
\mciteBstWouldAddEndPuncttrue
\mciteSetBstMidEndSepPunct{\mcitedefaultmidpunct}
{\mcitedefaultendpunct}{\mcitedefaultseppunct}\relax
\EndOfBibitem
\bibitem[Nguyen \latin{et~al.}(2021)Nguyen, Lombeck, Schwarz, Heizmann,
  Adamski, Lee, Britton, Holdcroft, Vierrath, and
  Breitwieser]{nguyen2021hydrocarbon}
Nguyen,~H.; Lombeck,~F.; Schwarz,~C.; Heizmann,~P.~A.; Adamski,~M.; Lee,~H.-F.;
  Britton,~B.; Holdcroft,~S.; Vierrath,~S.; Breitwieser,~M. Hydrocarbon-based
  Pemion™ proton exchange membrane fuel cells with state-of-the-art
  performance. \emph{Sustain. Energ. Fuels} \textbf{2021}, \emph{5},
  3687--3699\relax
\mciteBstWouldAddEndPuncttrue
\mciteSetBstMidEndSepPunct{\mcitedefaultmidpunct}
{\mcitedefaultendpunct}{\mcitedefaultseppunct}\relax
\EndOfBibitem
\bibitem[Nguyen \latin{et~al.}(2022)Nguyen, Klose, Metzler, Vierrath, and
  Breitwieser]{nguyen2022fully}
Nguyen,~H.; Klose,~C.; Metzler,~L.; Vierrath,~S.; Breitwieser,~M. Fully
  hydrocarbon membrane electrode assemblies for proton exchange membrane fuel
  cells and electrolyzers: An engineering perspective. \emph{Adv. Energ.
  Mater.} \textbf{2022}, \emph{12}, 2103559\relax
\mciteBstWouldAddEndPuncttrue
\mciteSetBstMidEndSepPunct{\mcitedefaultmidpunct}
{\mcitedefaultendpunct}{\mcitedefaultseppunct}\relax
\EndOfBibitem
\bibitem[Mirfarsi \latin{et~al.}(2024)Mirfarsi, Kumar, Jeong, Adamski,
  McDermid, Britton, and Kjeang]{mirfarsi2024high}
Mirfarsi,~S.~H.; Kumar,~A.; Jeong,~J.; Adamski,~M.; McDermid,~S.; Britton,~B.;
  Kjeang,~E. High-temperature stability of hydrocarbon-based
  Pemion{\textregistered} proton exchange membranes: A thermo-mechanical
  stability study. \emph{Int. J. Hydrogen Energy} \textbf{2024}, \emph{50},
  1507--1522\relax
\mciteBstWouldAddEndPuncttrue
\mciteSetBstMidEndSepPunct{\mcitedefaultmidpunct}
{\mcitedefaultendpunct}{\mcitedefaultseppunct}\relax
\EndOfBibitem
\bibitem[Ashcraft \latin{et~al.}(2010)Ashcraft, Argun, and
  Hammond]{ashcraft2010structure}
Ashcraft,~J.~N.; Argun,~A.~A.; Hammond,~P.~T. Structure-property studies of
  highly conductive layer-by-layer assembled membranes for fuel cell PEM
  applications. \emph{J. Mater. Chem.} \textbf{2010}, \emph{20},
  6250--6257\relax
\mciteBstWouldAddEndPuncttrue
\mciteSetBstMidEndSepPunct{\mcitedefaultmidpunct}
{\mcitedefaultendpunct}{\mcitedefaultseppunct}\relax
\EndOfBibitem
\bibitem[Afsar \latin{et~al.}(2019)Afsar, Ji, Wu, Shehzad, Ge, and
  Xu]{afsar2019sppo}
Afsar,~N.~U.; Ji,~W.; Wu,~B.; Shehzad,~M.~A.; Ge,~L.; Xu,~T. SPPO-based cation
  exchange membranes with a positively charged layer for cation fractionation.
  \emph{Desalination} \textbf{2019}, \emph{472}, 114145\relax
\mciteBstWouldAddEndPuncttrue
\mciteSetBstMidEndSepPunct{\mcitedefaultmidpunct}
{\mcitedefaultendpunct}{\mcitedefaultseppunct}\relax
\EndOfBibitem
\bibitem[Khan \latin{et~al.}(2022)Khan, Shanableh, Shahida, Lashari, Manzoor,
  and Fernandez]{khan2022speek}
Khan,~M.~I.; Shanableh,~A.; Shahida,~S.; Lashari,~M.~H.; Manzoor,~S.;
  Fernandez,~J. SPEEK and SPPO blended membranes for proton exchange membrane
  fuel cells. \emph{Membranes} \textbf{2022}, \emph{12}, 263\relax
\mciteBstWouldAddEndPuncttrue
\mciteSetBstMidEndSepPunct{\mcitedefaultmidpunct}
{\mcitedefaultendpunct}{\mcitedefaultseppunct}\relax
\EndOfBibitem
\bibitem[Yang \latin{et~al.}(2006)Yang, Gong, Guan, Zou, and
  Dai]{yang2006sulfonated}
Yang,~S.; Gong,~C.; Guan,~R.; Zou,~H.; Dai,~H. Sulfonated poly (phenylene
  oxide) membranes as promising materials for new proton exchange membranes.
  \emph{Polym. Adv. Technol.} \textbf{2006}, \emph{17}, 360--365\relax
\mciteBstWouldAddEndPuncttrue
\mciteSetBstMidEndSepPunct{\mcitedefaultmidpunct}
{\mcitedefaultendpunct}{\mcitedefaultseppunct}\relax
\EndOfBibitem
\bibitem[Xu \latin{et~al.}(2008)Xu, Wu, and Wu]{xu2008poly}
Xu,~T.; Wu,~D.; Wu,~L. Poly (2, 6-dimethyl-1, 4-phenylene oxide)(PPO)—a
  versatile starting polymer for proton conductive membranes (PCMs).
  \emph{Prog. Polym. Sci.} \textbf{2008}, \emph{33}, 894--915\relax
\mciteBstWouldAddEndPuncttrue
\mciteSetBstMidEndSepPunct{\mcitedefaultmidpunct}
{\mcitedefaultendpunct}{\mcitedefaultseppunct}\relax
\EndOfBibitem
\bibitem[Lee \latin{et~al.}(2016)Lee, Lee, Kim, Joo, Maurya, Choun, Lee, and
  Moon]{lee2016sppo}
Lee,~J.-H.; Lee,~J.-Y.; Kim,~J.-H.; Joo,~J.; Maurya,~S.; Choun,~M.; Lee,~J.;
  Moon,~S.-H. SPPO pore-filled composite membranes with electrically aligned
  ion channels via a lab-scale continuous caster for fuel cells: An optimal DC
  electric field strength-IEC relationship. \emph{J. Membr. Sci.}
  \textbf{2016}, \emph{501}, 15--23\relax
\mciteBstWouldAddEndPuncttrue
\mciteSetBstMidEndSepPunct{\mcitedefaultmidpunct}
{\mcitedefaultendpunct}{\mcitedefaultseppunct}\relax
\EndOfBibitem
\bibitem[Nagendra \latin{et~al.}(2023)Nagendra, Salman, Peter~Bloch, Daniel,
  Rizzo, Fittipaldi, and Guerra]{nagendra2023poly}
Nagendra,~B.; Salman,~S.; Peter~Bloch,~H.; Daniel,~C.; Rizzo,~P.;
  Fittipaldi,~R.; Guerra,~G. Poly (phenylene oxide) films with hydrophilic
  sulfonated amorphous phase and physically cross-linking hydrophobic
  crystalline phase. \emph{ACS Appl. Polym. Mater.} \textbf{2023}, \emph{5},
  3489--3498\relax
\mciteBstWouldAddEndPuncttrue
\mciteSetBstMidEndSepPunct{\mcitedefaultmidpunct}
{\mcitedefaultendpunct}{\mcitedefaultseppunct}\relax
\EndOfBibitem
\bibitem[Ionomr(2021)]{permionionomr}
Ionomr, \emph{Pemion$^{\rm \textregistered}$ Fuel Cell Offerings Proton
  Exchange Membranes \& Polymers}; 2021; PF1-HLF8-15-X, Document ID:
  FM-6027-B\relax
\mciteBstWouldAddEndPuncttrue
\mciteSetBstMidEndSepPunct{\mcitedefaultmidpunct}
{\mcitedefaultendpunct}{\mcitedefaultseppunct}\relax
\EndOfBibitem
\bibitem[inn()]{innochemtech}
Innochemtech Inc. \href{http://innochemtech.com/}{http://innochemtech.com/},
  accessed 2026-01-12\relax
\mciteBstWouldAddEndPuncttrue
\mciteSetBstMidEndSepPunct{\mcitedefaultmidpunct}
{\mcitedefaultendpunct}{\mcitedefaultseppunct}\relax
\EndOfBibitem
\bibitem[Kim \latin{et~al.}(2018)Kim, Chandrasekaran, Huan, Das, and
  Ramprasad]{Chiho:PG}
Kim,~C.; Chandrasekaran,~A.; Huan,~T.~D.; Das,~D.; Ramprasad,~R. Polymer
  Genome: A Data-Powered Polymer Informatics Platform for Property Predictions.
  \emph{J. Phys. Chem. C} \textbf{2018}, \emph{122}, 17575--17585\relax
\mciteBstWouldAddEndPuncttrue
\mciteSetBstMidEndSepPunct{\mcitedefaultmidpunct}
{\mcitedefaultendpunct}{\mcitedefaultseppunct}\relax
\EndOfBibitem
\bibitem[Tran \latin{et~al.}(2020)Tran, Kim, Chen, Chandrasekaran, Batra,
  Venkatram, Kamal, Lightstone, Gurnani, Shetty, Ramprasad, Laws, Shelton, and
  Ramprasad]{doan2020machine}
Tran,~H.; Kim,~C.; Chen,~L.; Chandrasekaran,~A.; Batra,~R.; Venkatram,~S.;
  Kamal,~D.; Lightstone,~J.~P.; Gurnani,~R.; Shetty,~P. \latin{et~al.}
  Machine-Learning Predictions of Polymer Properties with Polymer Genome.
  \emph{J. Appl. Phys.} \textbf{2020}, \emph{128}, 171104\relax
\mciteBstWouldAddEndPuncttrue
\mciteSetBstMidEndSepPunct{\mcitedefaultmidpunct}
{\mcitedefaultendpunct}{\mcitedefaultseppunct}\relax
\EndOfBibitem
\bibitem[Kim \latin{et~al.}(2023)Kim, Schroeder, and Jackson]{kim2023open}
Kim,~S.; Schroeder,~C.~M.; Jackson,~N.~E. Open macromolecular genome:
  Generative design of synthetically accessible polymers. \emph{ACS Polymers
  Au} \textbf{2023}, \emph{3}, 318--330\relax
\mciteBstWouldAddEndPuncttrue
\mciteSetBstMidEndSepPunct{\mcitedefaultmidpunct}
{\mcitedefaultendpunct}{\mcitedefaultseppunct}\relax
\EndOfBibitem
\bibitem[Ohno \latin{et~al.}(2023)Ohno, Hayashi, Zhang, Kaneko, and
  Yoshida]{ohno2023smipoly}
Ohno,~M.; Hayashi,~Y.; Zhang,~Q.; Kaneko,~Y.; Yoshida,~R. SMiPoly: generation
  of a synthesizable polymer virtual library using rule-based polymerization
  reactions. \emph{J. Chem. Infor. Model.} \textbf{2023}, \emph{63},
  5539--5548\relax
\mciteBstWouldAddEndPuncttrue
\mciteSetBstMidEndSepPunct{\mcitedefaultmidpunct}
{\mcitedefaultendpunct}{\mcitedefaultseppunct}\relax
\EndOfBibitem
\bibitem[Gurnani \latin{et~al.}(2024)Gurnani, Shukla, Kamal, Wu, Hao, Kuenneth,
  Aklujkar, Khomane, Daniels, Deshmukh, \latin{et~al.} others]{gurnani2024ai}
Gurnani,~R.; Shukla,~S.; Kamal,~D.; Wu,~C.; Hao,~J.; Kuenneth,~C.;
  Aklujkar,~P.; Khomane,~A.; Daniels,~R.; Deshmukh,~A.~A. \latin{et~al.}
  AI-assisted discovery of high-temperature dielectrics for energy storage.
  \emph{Nat. Commun.} \textbf{2024}, \emph{15}, 6107\relax
\mciteBstWouldAddEndPuncttrue
\mciteSetBstMidEndSepPunct{\mcitedefaultmidpunct}
{\mcitedefaultendpunct}{\mcitedefaultseppunct}\relax
\EndOfBibitem
\bibitem[Kern \latin{et~al.}(2025)Kern, Su, Gutekunst, and
  Ramprasad]{kern2025informatics}
Kern,~J.; Su,~Y.-L.; Gutekunst,~W.; Ramprasad,~R. An informatics framework for
  the design of sustainable, chemically recyclable, synthetically accessible,
  and durable polymers. \emph{npj Comput. Mater.} \textbf{2025}, \emph{11},
  182\relax
\mciteBstWouldAddEndPuncttrue
\mciteSetBstMidEndSepPunct{\mcitedefaultmidpunct}
{\mcitedefaultendpunct}{\mcitedefaultseppunct}\relax
\EndOfBibitem
\bibitem[Li(2005)]{li2005principles}
Li,~X. \emph{Principles of Fuel Cells}; CRC press, 2005\relax
\mciteBstWouldAddEndPuncttrue
\mciteSetBstMidEndSepPunct{\mcitedefaultmidpunct}
{\mcitedefaultendpunct}{\mcitedefaultseppunct}\relax
\EndOfBibitem
\bibitem[Jiao and Li(2011)Jiao, and Li]{jiao2011water}
Jiao,~K.; Li,~X. Water Transport in Polymer Electrolyte Membrane Fuel Cells.
  \emph{Prog. Energy Combust. Sci.} \textbf{2011}, \emph{37}, 221--291\relax
\mciteBstWouldAddEndPuncttrue
\mciteSetBstMidEndSepPunct{\mcitedefaultmidpunct}
{\mcitedefaultendpunct}{\mcitedefaultseppunct}\relax
\EndOfBibitem
\bibitem[Kusoglu and Weber(2017)Kusoglu, and Weber]{kusoglu2017new}
Kusoglu,~A.; Weber,~A.~Z. New Insights into Perfluorinated Sulfonic-Acid
  Ionomers. \emph{Chem. Rev.} \textbf{2017}, \emph{117}, 987--1104\relax
\mciteBstWouldAddEndPuncttrue
\mciteSetBstMidEndSepPunct{\mcitedefaultmidpunct}
{\mcitedefaultendpunct}{\mcitedefaultseppunct}\relax
\EndOfBibitem
\bibitem[Kusoglu and Weber(2012)Kusoglu, and Weber]{kusoglu2012water}
Kusoglu,~A.; Weber,~A.~Z. \emph{Polymers for Energy Storage and Delivery:
  Polyelectrolytes for Batteries and Fuel Cells}; ACS Publications, 2012;
  Chapter 11, pp 175--199\relax
\mciteBstWouldAddEndPuncttrue
\mciteSetBstMidEndSepPunct{\mcitedefaultmidpunct}
{\mcitedefaultendpunct}{\mcitedefaultseppunct}\relax
\EndOfBibitem
\bibitem[Hickner and Pivovar(2005)Hickner, and Pivovar]{hickner2005chemical}
Hickner,~M.; Pivovar,~B. The Chemical and Structural Nature of Proton Exchange
  Membrane Fuel Cell Properties. \emph{Fuel cells} \textbf{2005}, \emph{5},
  213--229\relax
\mciteBstWouldAddEndPuncttrue
\mciteSetBstMidEndSepPunct{\mcitedefaultmidpunct}
{\mcitedefaultendpunct}{\mcitedefaultseppunct}\relax
\EndOfBibitem
\bibitem[Zhu \latin{et~al.}(2022)Zhu, Li, Liu, He, Wang, and
  Lei]{zhu2022recent}
Zhu,~L.-Y.; Li,~Y.-C.; Liu,~J.; He,~J.; Wang,~L.-Y.; Lei,~J.-D. Recent
  Developments in High-Performance Nafion Membranes for Hydrogen Fuel Cells
  Applications. \emph{Pet. Sci.} \textbf{2022}, \emph{19}, 1371\relax
\mciteBstWouldAddEndPuncttrue
\mciteSetBstMidEndSepPunct{\mcitedefaultmidpunct}
{\mcitedefaultendpunct}{\mcitedefaultseppunct}\relax
\EndOfBibitem
\bibitem[Feng \latin{et~al.}(2018)Feng, Kondo, Kaseyama, Nakazawa, Kikuchi,
  Selyanchyn, Fujikawa, Christiani, Sasaki, and
  Nishihara]{feng2018characterization}
Feng,~S.; Kondo,~S.; Kaseyama,~T.; Nakazawa,~T.; Kikuchi,~T.; Selyanchyn,~R.;
  Fujikawa,~S.; Christiani,~L.; Sasaki,~K.; Nishihara,~M. Characterization of
  Polymer-Polymer Type Charge-Transfer (CT) Blend Membranes for Fuel Cell
  Application. \emph{Data Br.} \textbf{2018}, \emph{18}, 22--29\relax
\mciteBstWouldAddEndPuncttrue
\mciteSetBstMidEndSepPunct{\mcitedefaultmidpunct}
{\mcitedefaultendpunct}{\mcitedefaultseppunct}\relax
\EndOfBibitem
\bibitem[Zhang \latin{et~al.}(2018)Zhang, Miyake, Akiyama, Shimizu, and
  Miyatake]{zhang2018sulfonated}
Zhang,~Y.; Miyake,~J.; Akiyama,~R.; Shimizu,~R.; Miyatake,~K. Sulfonated
  Phenylene/Quinquephenylene/Perfluoroalkylene Terpolymers as Proton Exchange
  Membranes for Fuel Cells. \emph{ACS Appl. Energy Mater.} \textbf{2018},
  \emph{1}, 1008--1015\relax
\mciteBstWouldAddEndPuncttrue
\mciteSetBstMidEndSepPunct{\mcitedefaultmidpunct}
{\mcitedefaultendpunct}{\mcitedefaultseppunct}\relax
\EndOfBibitem
\bibitem[Peng \latin{et~al.}(2017)Peng, Lai, and Liu]{peng2017preparation}
Peng,~K.-J.; Lai,~J.-Y.; Liu,~Y.-L. Preparation of Poly (Styrenesulfonic Acid)
  Grafted Nafion with a Nafion-Initiated Atom Transfer Radical Polymerization
  for Proton Exchange Membranes. \emph{RSC Adv.} \textbf{2017}, \emph{7},
  37255--37260\relax
\mciteBstWouldAddEndPuncttrue
\mciteSetBstMidEndSepPunct{\mcitedefaultmidpunct}
{\mcitedefaultendpunct}{\mcitedefaultseppunct}\relax
\EndOfBibitem
\bibitem[Si \latin{et~al.}(2012)Si, Dong, Wycisk, and Litt]{si2012synthesis}
Si,~K.; Dong,~D.; Wycisk,~R.; Litt,~M. Synthesis and Characterization of Poly
  (Para-Phenylene Disulfonic Acid), its Copolymers and Their n-Alkylbenzene
  Grafts as Proton Exchange Membranes: High Conductivity at Low Relative
  Humidity. \emph{J. Mater. Chem.} \textbf{2012}, \emph{22}, 20907--20917\relax
\mciteBstWouldAddEndPuncttrue
\mciteSetBstMidEndSepPunct{\mcitedefaultmidpunct}
{\mcitedefaultendpunct}{\mcitedefaultseppunct}\relax
\EndOfBibitem
\bibitem[Wang \latin{et~al.}(2012)Wang, Shin, Lee, Kang, Robertson, Lee, and
  Guiver]{wang2012clustered}
Wang,~C.; Shin,~D.~W.; Lee,~S.~Y.; Kang,~N.~R.; Robertson,~G.~P.; Lee,~Y.~M.;
  Guiver,~M.~D. A Clustered Sulfonated Poly (Ether Sulfone) Based on a New
  Fluorene-Based Bisphenol Monomer. \emph{J. Mater. Chem.} \textbf{2012},
  \emph{22}, 25093--25101\relax
\mciteBstWouldAddEndPuncttrue
\mciteSetBstMidEndSepPunct{\mcitedefaultmidpunct}
{\mcitedefaultendpunct}{\mcitedefaultseppunct}\relax
\EndOfBibitem
\bibitem[Silva \latin{et~al.}(2004)Silva, De~Francesco, and
  Pozio]{silva2004tangential}
Silva,~R.; De~Francesco,~M.; Pozio,~A. Tangential and normal conductivities of
  Nafion{\textregistered} membranes used in polymer electrolyte fuel cells.
  \emph{J. Power Sources} \textbf{2004}, \emph{134}, 18--26\relax
\mciteBstWouldAddEndPuncttrue
\mciteSetBstMidEndSepPunct{\mcitedefaultmidpunct}
{\mcitedefaultendpunct}{\mcitedefaultseppunct}\relax
\EndOfBibitem
\bibitem[Roberti \latin{et~al.}(2010)Roberti, Carlotti, Cinelli, Onori,
  Donnadio, Narducci, Casciola, and Sganappa]{roberti2010measurement}
Roberti,~E.; Carlotti,~G.; Cinelli,~S.; Onori,~G.; Donnadio,~A.; Narducci,~R.;
  Casciola,~M.; Sganappa,~M. Measurement of the Young's Modulus of Nafion
  Membranes by Brillouin Light Scattering. \emph{J. Power Sources}
  \textbf{2010}, \emph{195}, 7761--7764\relax
\mciteBstWouldAddEndPuncttrue
\mciteSetBstMidEndSepPunct{\mcitedefaultmidpunct}
{\mcitedefaultendpunct}{\mcitedefaultseppunct}\relax
\EndOfBibitem
\bibitem[Caire \latin{et~al.}(2016)Caire, Vandiver, Pandey, Herring, and
  Liberatore]{caire2016accelerated}
Caire,~B.~R.; Vandiver,~M.~A.; Pandey,~T.~P.; Herring,~A.~M.; Liberatore,~M.~W.
  Accelerated Mechanical Degradation of Anion Exchange Membranes via Hydration
  Cycling. \emph{J. Electrochem. Soc.} \textbf{2016}, \emph{163}, H964\relax
\mciteBstWouldAddEndPuncttrue
\mciteSetBstMidEndSepPunct{\mcitedefaultmidpunct}
{\mcitedefaultendpunct}{\mcitedefaultseppunct}\relax
\EndOfBibitem
\bibitem[Jung and Kim(2012)Jung, and Kim]{jung2012role}
Jung,~H.-Y.; Kim,~J.~W. Role of the Glass Transition Temperature of Nafion 117
  Membrane in the Preparation of the Membrane Electrode Assembly in a Direct
  Methanol Fuel Cell (DMFC). \emph{Int. J. Hydrog. Energy} \textbf{2012},
  \emph{37}, 12580--12585\relax
\mciteBstWouldAddEndPuncttrue
\mciteSetBstMidEndSepPunct{\mcitedefaultmidpunct}
{\mcitedefaultendpunct}{\mcitedefaultseppunct}\relax
\EndOfBibitem
\bibitem[Meyer \latin{et~al.}(2017)Meyer, Mansor, Iacoviello, Cullen, Jervis,
  Finegan, Tan, Bailey, Shearing, and Brett]{meyer2017investigation}
Meyer,~Q.; Mansor,~N.; Iacoviello,~F.; Cullen,~P.; Jervis,~R.; Finegan,~D.;
  Tan,~C.; Bailey,~J.; Shearing,~P.; Brett,~D. Investigation of Hot Pressed
  Polymer Electrolyte Fuel Cell Assemblies via X-Ray Computed Tomography.
  \emph{Electrochim. Acta} \textbf{2017}, \emph{242}, 125--136\relax
\mciteBstWouldAddEndPuncttrue
\mciteSetBstMidEndSepPunct{\mcitedefaultmidpunct}
{\mcitedefaultendpunct}{\mcitedefaultseppunct}\relax
\EndOfBibitem
\bibitem[Lage \latin{et~al.}(2004)Lage, Delgado, and Kawano]{lage2004thermal}
Lage,~L.; Delgado,~P.; Kawano,~Y. Thermal Stability and Decomposition of
  Nafion\textsuperscript{\textregistered} Membranes with Different Cations.
  \emph{J. Therm. Anal. Calorim.} \textbf{2004}, \emph{75}, 521--530\relax
\mciteBstWouldAddEndPuncttrue
\mciteSetBstMidEndSepPunct{\mcitedefaultmidpunct}
{\mcitedefaultendpunct}{\mcitedefaultseppunct}\relax
\EndOfBibitem
\bibitem[Samms \latin{et~al.}(1996)Samms, Wasmus, and
  Savinell]{samms1996thermal}
Samms,~S.; Wasmus,~S.; Savinell,~R. Thermal Stability of
  Nafion\textsuperscript{\textregistered} in Simulated Fuel Cell Environments.
  \emph{J. Electrochem. Soc.} \textbf{1996}, \emph{143}, 1498\relax
\mciteBstWouldAddEndPuncttrue
\mciteSetBstMidEndSepPunct{\mcitedefaultmidpunct}
{\mcitedefaultendpunct}{\mcitedefaultseppunct}\relax
\EndOfBibitem
\bibitem[Mukaddam \latin{et~al.}(2016)Mukaddam, Litwiller, and
  Pinnau]{mukaddam2016gas}
Mukaddam,~M.; Litwiller,~E.; Pinnau,~I. Gas Sorption, Diffusion, and Permeation
  in Nafion. \emph{Macromolecules} \textbf{2016}, \emph{49}, 280--286\relax
\mciteBstWouldAddEndPuncttrue
\mciteSetBstMidEndSepPunct{\mcitedefaultmidpunct}
{\mcitedefaultendpunct}{\mcitedefaultseppunct}\relax
\EndOfBibitem
\bibitem[Fan \latin{et~al.}(2014)Fan, Tongren, and Cornelius]{fan2014role}
Fan,~Y.; Tongren,~D.; Cornelius,~C.~J. The Role of a Metal Ion within Nafion
  upon its Physical and Gas Transport Properties. \emph{Eur. Polym. J.}
  \textbf{2014}, \emph{50}, 271--278\relax
\mciteBstWouldAddEndPuncttrue
\mciteSetBstMidEndSepPunct{\mcitedefaultmidpunct}
{\mcitedefaultendpunct}{\mcitedefaultseppunct}\relax
\EndOfBibitem
\bibitem[Chiou and Paul(1988)Chiou, and Paul]{chiou1988gas}
Chiou,~J.~S.; Paul,~D.~R. Gas Permeation in a Dry Nafion Membrane. \emph{Ind.
  Eng. Chem. Res.} \textbf{1988}, \emph{27}, 2161--2164\relax
\mciteBstWouldAddEndPuncttrue
\mciteSetBstMidEndSepPunct{\mcitedefaultmidpunct}
{\mcitedefaultendpunct}{\mcitedefaultseppunct}\relax
\EndOfBibitem
\bibitem[Lowry and Mauritz(1980)Lowry, and Mauritz]{lowry1980investigation}
Lowry,~S.; Mauritz,~K. An Investigation of Ionic Hydration Effects in
  Perfluorosulfonate Ionomers by Fourier Transform Infrared Spectroscopy.
  \emph{J. Am. Chem. Soc.} \textbf{1980}, \emph{102}, 4665--4667\relax
\mciteBstWouldAddEndPuncttrue
\mciteSetBstMidEndSepPunct{\mcitedefaultmidpunct}
{\mcitedefaultendpunct}{\mcitedefaultseppunct}\relax
\EndOfBibitem
\bibitem[Huan \latin{et~al.}(2016)Huan, Mannodi-Kanakkithodi, Kim, Sharma,
  Pilania, and Ramprasad]{Huan:Data}
Huan,~T.~D.; Mannodi-Kanakkithodi,~A.; Kim,~C.; Sharma,~V.; Pilania,~G.;
  Ramprasad,~R. A Polymer Dataset for Accelerated Property Prediction and
  Design. \emph{Sci. Data} \textbf{2016}, \emph{3}, 160012\relax
\mciteBstWouldAddEndPuncttrue
\mciteSetBstMidEndSepPunct{\mcitedefaultmidpunct}
{\mcitedefaultendpunct}{\mcitedefaultseppunct}\relax
\EndOfBibitem
\bibitem[Kamal \latin{et~al.}(2021)Kamal, Tran, Kim, Wang, Chen, Cao, Joseph,
  and Ramprasad]{kamal2021novel}
Kamal,~D.; Tran,~H.; Kim,~C.; Wang,~Y.; Chen,~L.; Cao,~Y.; Joseph,~V.~R.;
  Ramprasad,~R. Novel High Voltage Polymer Insulators using Computational and
  Data-Driven Techniques. \emph{J. Chem. Phys.} \textbf{2021}, \emph{154},
  174906\relax
\mciteBstWouldAddEndPuncttrue
\mciteSetBstMidEndSepPunct{\mcitedefaultmidpunct}
{\mcitedefaultendpunct}{\mcitedefaultseppunct}\relax
\EndOfBibitem
\bibitem[Kamal \latin{et~al.}(2020)Kamal, Wang, Tran, Chen, Li, Wu, Nasreen,
  Cao, and Ramprasad]{kamal2020computable}
Kamal,~D.; Wang,~Y.; Tran,~H.~D.; Chen,~L.; Li,~Z.; Wu,~C.; Nasreen,~S.;
  Cao,~Y.; Ramprasad,~R. Computable Bulk and Interfacial Electronic Structure
  Features as Proxies for Dielectric Breakdown of Polymers. \emph{ACS Appl.
  Mater. Interfaces} \textbf{2020}, \emph{12}, 37182--37187\relax
\mciteBstWouldAddEndPuncttrue
\mciteSetBstMidEndSepPunct{\mcitedefaultmidpunct}
{\mcitedefaultendpunct}{\mcitedefaultseppunct}\relax
\EndOfBibitem
\bibitem[Weininger(1988)]{smiles}
Weininger,~D. SMILES, a Chemical Language and Information System. 1.
  Introduction to Methodology and Encoding Rules. \emph{J. Chem. Inf. Comput.
  Sci.} \textbf{1988}, \emph{28}, 31--36\relax
\mciteBstWouldAddEndPuncttrue
\mciteSetBstMidEndSepPunct{\mcitedefaultmidpunct}
{\mcitedefaultendpunct}{\mcitedefaultseppunct}\relax
\EndOfBibitem
\bibitem[Pilania \latin{et~al.}(2013)Pilania, Wang, Jiang, Rajasekaran, and
  Ramprasad]{Pilania_SR}
Pilania,~G.; Wang,~C.; Jiang,~X.; Rajasekaran,~S.; Ramprasad,~R. Accelerating
  Materials Property Predictions using Machine Learning. \emph{Sci. Rep.}
  \textbf{2013}, \emph{3}, 2810\relax
\mciteBstWouldAddEndPuncttrue
\mciteSetBstMidEndSepPunct{\mcitedefaultmidpunct}
{\mcitedefaultendpunct}{\mcitedefaultseppunct}\relax
\EndOfBibitem
\bibitem[Huan \latin{et~al.}(2015)Huan, Mannodi-Kanakkithodi, and
  Ramprasad]{Huan:design}
Huan,~T.~D.; Mannodi-Kanakkithodi,~A.; Ramprasad,~R. Accelerated Materials
  Property Predictions and Design using Motif-Based Fingerprints. \emph{Phys.
  Rev. B} \textbf{2015}, \emph{92}, 014106\relax
\mciteBstWouldAddEndPuncttrue
\mciteSetBstMidEndSepPunct{\mcitedefaultmidpunct}
{\mcitedefaultendpunct}{\mcitedefaultseppunct}\relax
\EndOfBibitem
\bibitem[Mannodi-Kanakkithodi \latin{et~al.}(2016)Mannodi-Kanakkithodi,
  Pilania, Huan, Lookman, and Ramprasad]{Arun:design}
Mannodi-Kanakkithodi,~A.; Pilania,~G.; Huan,~T.~D.; Lookman,~T.; Ramprasad,~R.
  Machine Learning Strategy for the Accelerated Design of Polymer Dielectrics.
  \emph{Sci. Rep.} \textbf{2016}, \emph{6}, 20952\relax
\mciteBstWouldAddEndPuncttrue
\mciteSetBstMidEndSepPunct{\mcitedefaultmidpunct}
{\mcitedefaultendpunct}{\mcitedefaultseppunct}\relax
\EndOfBibitem
\bibitem[Rasmussen and Williams(2006)Rasmussen, and Williams]{GPRBook}
Rasmussen,~C.~E., Williams,~C. K.~I., Eds. \emph{Gaussian Processes for Machine
  Learning}; The MIT Press: Cambridge, MA, 2006\relax
\mciteBstWouldAddEndPuncttrue
\mciteSetBstMidEndSepPunct{\mcitedefaultmidpunct}
{\mcitedefaultendpunct}{\mcitedefaultseppunct}\relax
\EndOfBibitem
\bibitem[Williams and Rasmussen(1995)Williams, and Rasmussen]{GPR95}
Williams,~C. K.~I.; Rasmussen,~C.~E. In \emph{Advances in Neural Information
  Processing Systems 8}; Touretzky,~D.~S., Mozer,~M.~C., Hasselmo,~M.~E., Eds.;
  MIT Press, 1995\relax
\mciteBstWouldAddEndPuncttrue
\mciteSetBstMidEndSepPunct{\mcitedefaultmidpunct}
{\mcitedefaultendpunct}{\mcitedefaultseppunct}\relax
\EndOfBibitem
\bibitem[Patra \latin{et~al.}(2020)Patra, Batra, Chandrasekaran, Kim, Huan, and
  Ramprasad]{PATRA2020109286}
Patra,~A.; Batra,~R.; Chandrasekaran,~A.; Kim,~C.; Huan,~T.~D.; Ramprasad,~R. A
  multi-fidelity information-fusion approach to machine learn and predict
  polymer bandgap. \emph{Comput. Mater. Sci.} \textbf{2020}, \emph{172},
  109286\relax
\mciteBstWouldAddEndPuncttrue
\mciteSetBstMidEndSepPunct{\mcitedefaultmidpunct}
{\mcitedefaultendpunct}{\mcitedefaultseppunct}\relax
\EndOfBibitem
\bibitem[Kuenneth \latin{et~al.}(2021)Kuenneth, Rajan, Tran, Chen, Kim, and
  Ramprasad]{kuenneth2021polymer}
Kuenneth,~C.; Rajan,~A.~C.; Tran,~H.; Chen,~L.; Kim,~C.; Ramprasad,~R. Polymer
  Informatics with Multi-Task Learning. \emph{Patterns} \textbf{2021},
  \emph{2}, 100238\relax
\mciteBstWouldAddEndPuncttrue
\mciteSetBstMidEndSepPunct{\mcitedefaultmidpunct}
{\mcitedefaultendpunct}{\mcitedefaultseppunct}\relax
\EndOfBibitem
\bibitem[Tran \latin{et~al.}(2025)Tran, Kim, Gurnani, Hvidsten, DeSimpliciis,
  Ramprasad, Gadelrab, Tuffile, Molinari, Kitchaev, \latin{et~al.}
  others]{tran2025polymer}
Tran,~H.; Kim,~C.; Gurnani,~R.; Hvidsten,~O.; DeSimpliciis,~J.; Ramprasad,~R.;
  Gadelrab,~K.; Tuffile,~C.; Molinari,~N.; Kitchaev,~D. \latin{et~al.}  Polymer
  composites informatics for flammability, thermal, mechanical and electrical
  property predictions. \emph{Polym. Chem.} \textbf{2025}, \emph{16},
  3459--3467\relax
\mciteBstWouldAddEndPuncttrue
\mciteSetBstMidEndSepPunct{\mcitedefaultmidpunct}
{\mcitedefaultendpunct}{\mcitedefaultseppunct}\relax
\EndOfBibitem
\bibitem[Shukla \latin{et~al.}(2026)Shukla, Kim, Gurnani, Ramprasad, and
  Mahmood]{akhlak:rxnchainer}
Shukla,~S.~S.; Kim,~C.; Gurnani,~R.; Ramprasad,~R.; Mahmood,~A. Stalking the
  Synthetically-Accessible Polymer Universe: Virtual Synthesis, Retrosynthesis
  and Beyond. \emph{Preprints.org} \textbf{2026}, Preprint, url
  \texttt{https://doi.org/10.20944/preprints202605.0977.v1}\relax
\mciteBstWouldAddEndPuncttrue
\mciteSetBstMidEndSepPunct{\mcitedefaultmidpunct}
{\mcitedefaultendpunct}{\mcitedefaultseppunct}\relax
\EndOfBibitem
\bibitem[TSC()]{TSCA}
TSCA Chemical Substance Inventory.
  \href{https://www.epa.gov/tsca-inventory}{https://www.epa.gov/tsca-inventory},
  accessed 2026-01-17\relax
\mciteBstWouldAddEndPuncttrue
\mciteSetBstMidEndSepPunct{\mcitedefaultmidpunct}
{\mcitedefaultendpunct}{\mcitedefaultseppunct}\relax
\EndOfBibitem
\bibitem[Tingle \latin{et~al.}(2023)Tingle, Tang, Castanon, Gutierrez,
  Khurelbaatar, Dandarchuluun, Moroz, and Irwin]{tingle2023zinc}
Tingle,~B.~I.; Tang,~K.~G.; Castanon,~M.; Gutierrez,~J.~J.; Khurelbaatar,~M.;
  Dandarchuluun,~C.; Moroz,~Y.~S.; Irwin,~J.~J. ZINC-22 -- A free
  multi-billion-scale database of tangible compounds for ligand discovery.
  \emph{J. Chem. Inf. Model.} \textbf{2023}, \emph{63}, 1166--1176\relax
\mciteBstWouldAddEndPuncttrue
\mciteSetBstMidEndSepPunct{\mcitedefaultmidpunct}
{\mcitedefaultendpunct}{\mcitedefaultseppunct}\relax
\EndOfBibitem
\bibitem[Mayr \latin{et~al.}(2018)Mayr, Klambauer, Unterthiner, Steijaert,
  Wegner, Ceulemans, Clevert, and Hochreiter]{mayr2018large}
Mayr,~A.; Klambauer,~G.; Unterthiner,~T.; Steijaert,~M.; Wegner,~J.~K.;
  Ceulemans,~H.; Clevert,~D.-A.; Hochreiter,~S. Large-scale comparison of
  machine learning methods for drug target prediction on ChEMBL. \emph{Chem.
  Sci.} \textbf{2018}, \emph{9}, 5441--5451\relax
\mciteBstWouldAddEndPuncttrue
\mciteSetBstMidEndSepPunct{\mcitedefaultmidpunct}
{\mcitedefaultendpunct}{\mcitedefaultseppunct}\relax
\EndOfBibitem
\bibitem[eMo()]{eMolecules}
eMolecules. \href{https://www.emolecules.com/}{https://www.emolecules.com/},
  accessed 2026-01-15\relax
\mciteBstWouldAddEndPuncttrue
\mciteSetBstMidEndSepPunct{\mcitedefaultmidpunct}
{\mcitedefaultendpunct}{\mcitedefaultseppunct}\relax
\EndOfBibitem
\bibitem[{Matmerize, Inc.}()]{polymrize_url}
{Matmerize, Inc.}, PolymRize.
  \href{https://polymrize.matmerize.com/}{https://polymrize.matmerize.com/}\relax
\mciteBstWouldAddEndPuncttrue
\mciteSetBstMidEndSepPunct{\mcitedefaultmidpunct}
{\mcitedefaultendpunct}{\mcitedefaultseppunct}\relax
\EndOfBibitem
\bibitem[McGrath \latin{et~al.}(2001)McGrath, Hickner, Wang, and
  Kim]{McGrath2001ionconducting}
McGrath,~E.,~James; Hickner,~M.; Wang,~F.; Kim,~Y.-S. Ion-conducting sulfonated
  polymeric materials. Patent EP1327278B1, 2001\relax
\mciteBstWouldAddEndPuncttrue
\mciteSetBstMidEndSepPunct{\mcitedefaultmidpunct}
{\mcitedefaultendpunct}{\mcitedefaultseppunct}\relax
\EndOfBibitem
\bibitem[Mirfarsi \latin{et~al.}(2025)Mirfarsi, Kumar, Jeong, Brown, Adamski,
  Jones, Mcdermid, Britton, and Kjeang]{mirfarsi2025mechanical}
Mirfarsi,~S.~H.; Kumar,~A.; Jeong,~J.; Brown,~E.; Adamski,~M.; Jones,~S.;
  Mcdermid,~S.; Britton,~B.; Kjeang,~E. Mechanical durability of reinforced
  sulfo-phenylated polyphenylene-based proton exchange membranes: Impacts of
  ion exchange capacity and reinforcement thickness. \emph{J.Power Sources}
  \textbf{2025}, \emph{630}, 236137\relax
\mciteBstWouldAddEndPuncttrue
\mciteSetBstMidEndSepPunct{\mcitedefaultmidpunct}
{\mcitedefaultendpunct}{\mcitedefaultseppunct}\relax
\EndOfBibitem
\bibitem[Mirfarsi \latin{et~al.}(2023)Mirfarsi, Kumar, Jeong, Adamski,
  McDermid, Britton, and Kjeang]{mirfarsi2023thermo}
Mirfarsi,~S.~H.; Kumar,~A.; Jeong,~J.; Adamski,~M.; McDermid,~S.; Britton,~B.;
  Kjeang,~E. Thermo-Mechanical Stability of Hydrocarbon-Based
  Pemion{\textregistered} Proton Exchange Membranes. Electrochemical Society
  Meeting Abstracts 244. 2023; pp 1903--1903\relax
\mciteBstWouldAddEndPuncttrue
\mciteSetBstMidEndSepPunct{\mcitedefaultmidpunct}
{\mcitedefaultendpunct}{\mcitedefaultseppunct}\relax
\EndOfBibitem
\bibitem[Petreanu \latin{et~al.}(2012)Petreanu, Ebrasu, Sisu, and
  Varlam]{petreanu2012thermal}
Petreanu,~I.; Ebrasu,~D.; Sisu,~C.; Varlam,~M. Thermal analysis of sulfonated
  polymers tested as polymer electrolyte membrane for PEM fuel cells. \emph{J.
  Therm. Anal. Calorim.} \textbf{2012}, \emph{110}, 335--339\relax
\mciteBstWouldAddEndPuncttrue
\mciteSetBstMidEndSepPunct{\mcitedefaultmidpunct}
{\mcitedefaultendpunct}{\mcitedefaultseppunct}\relax
\EndOfBibitem
\bibitem[{Ionomr Innovations Inc.}(2021)]{pemion_td}
{Ionomr Innovations Inc.}, \emph{Pemion$^{\rm \textregistered}$ Safety data
  sheet (DSC)}; 2021; FM-7009-H, Document ID: FM-7009-H\relax
\mciteBstWouldAddEndPuncttrue
\mciteSetBstMidEndSepPunct{\mcitedefaultmidpunct}
{\mcitedefaultendpunct}{\mcitedefaultseppunct}\relax
\EndOfBibitem
\bibitem[Garc{\'\i}a-Salaberri(2023)]{garcia2023proton}
Garc{\'\i}a-Salaberri,~P.~A. Proton exchange membranes for polymer electrolyte
  fuel cells: An analysis of perfluorosulfonic acid and aromatic hydrocarbon
  ionomers. \emph{Sustain. Mater. Technol.} \textbf{2023}, \emph{38},
  e00727\relax
\mciteBstWouldAddEndPuncttrue
\mciteSetBstMidEndSepPunct{\mcitedefaultmidpunct}
{\mcitedefaultendpunct}{\mcitedefaultseppunct}\relax
\EndOfBibitem
\bibitem[Kruczek(2001)]{kruczek2001gas}
Kruczek,~B. \emph{Polyphenylene Oxide and Modified Polyphenylene Oxide
  Membranes: Gas, Vapor and Liquid Separation}; Springer, 2001; pp
  61--104\relax
\mciteBstWouldAddEndPuncttrue
\mciteSetBstMidEndSepPunct{\mcitedefaultmidpunct}
{\mcitedefaultendpunct}{\mcitedefaultseppunct}\relax
\EndOfBibitem
\bibitem[Shiino \latin{et~al.}(2020)Shiino, Otomo, Yamada, Arima, Hiroi,
  Takata, Miyake, and Miyatake]{shiino2020structural}
Shiino,~K.; Otomo,~T.; Yamada,~T.; Arima,~H.; Hiroi,~K.; Takata,~S.;
  Miyake,~J.; Miyatake,~K. Structural investigation of sulfonated polyphenylene
  ionomers for the design of better performing proton-conductive membranes.
  \emph{ACS Appl. Polym. Mater.} \textbf{2020}, \emph{2}, 5558--5565\relax
\mciteBstWouldAddEndPuncttrue
\mciteSetBstMidEndSepPunct{\mcitedefaultmidpunct}
{\mcitedefaultendpunct}{\mcitedefaultseppunct}\relax
\EndOfBibitem
\bibitem[Cable \latin{et~al.}(1995)Cable, Mauritz, and Moore]{cable1995effects}
Cable,~K.~M.; Mauritz,~K.~A.; Moore,~R.~B. Effects of Hydrophilic and
  Hydrophobic Counterions on the Coulombic interactions in Perfluorosulfonate
  Ionomers. \emph{J. Polym. Sci. B Polym. Phys.} \textbf{1995}, \emph{33},
  1065--1072\relax
\mciteBstWouldAddEndPuncttrue
\mciteSetBstMidEndSepPunct{\mcitedefaultmidpunct}
{\mcitedefaultendpunct}{\mcitedefaultseppunct}\relax
\EndOfBibitem
\bibitem[Korzeniewski \latin{et~al.}(2008)Korzeniewski, Adams, and
  Liu]{korzeniewski2008responses}
Korzeniewski,~C.; Adams,~E.; Liu,~D. Responses of Hydrophobic and Hydrophilic
  Groups in Nafion Differentiated by Least Squares Modeling of Infrared Spectra
  Recorded During Thin Film Hydration. \emph{Appl. Spectrosc.} \textbf{2008},
  \emph{62}, 634--639\relax
\mciteBstWouldAddEndPuncttrue
\mciteSetBstMidEndSepPunct{\mcitedefaultmidpunct}
{\mcitedefaultendpunct}{\mcitedefaultseppunct}\relax
\EndOfBibitem
\bibitem[Liu \latin{et~al.}(2022)Liu, Lu, Ning, Li, Chen, and Hu]{liu2022comb}
Liu,~L.; Lu,~Y.; Ning,~C.; Li,~N.; Chen,~S.; Hu,~Z. Comb-Shaped Sulfonated Poly
  (aryl ether sulfone) Proton Exchange Membrane for Fuel Cell Applications.
  \emph{Int. J. Hydrog. Energy} \textbf{2022}, \emph{47}, 16249--16261\relax
\mciteBstWouldAddEndPuncttrue
\mciteSetBstMidEndSepPunct{\mcitedefaultmidpunct}
{\mcitedefaultendpunct}{\mcitedefaultseppunct}\relax
\EndOfBibitem
\bibitem[Sutradhar \latin{et~al.}(2025)Sutradhar, Bae, Song, Joo, Na, Lee, Kim,
  and Jang]{sutradhar2025synthesis}
Sutradhar,~S.~C.; Bae,~W.; Song,~S.; Joo,~K.; Na,~H.; Lee,~J.; Kim,~W.;
  Jang,~H. Synthesis and characterization of fluoro sulfonyl imide-based poly
  (benzoyl diphenyl benzene) membranes for proton exchange membrane fuel cells.
  \emph{Fuel} \textbf{2025}, \emph{390}, 134741\relax
\mciteBstWouldAddEndPuncttrue
\mciteSetBstMidEndSepPunct{\mcitedefaultmidpunct}
{\mcitedefaultendpunct}{\mcitedefaultseppunct}\relax
\EndOfBibitem
\bibitem[Tsakanika \latin{et~al.}(2026)Tsakanika, Stergiou, Pantazidis, and
  Sakellariou]{tsakanika2026single}
Tsakanika,~M.; Stergiou,~A.; Pantazidis,~C.; Sakellariou,~G. Single-ion Solid
  Polymer Electrolytes by Design: Chemistries, Architectures and Functional
  Trade-Offs. \emph{J. Mater. Chem. A} \textbf{2026}, DOI:
  10.1039/D6TA01035K\relax
\mciteBstWouldAddEndPuncttrue
\mciteSetBstMidEndSepPunct{\mcitedefaultmidpunct}
{\mcitedefaultendpunct}{\mcitedefaultseppunct}\relax
\EndOfBibitem
\bibitem[Kim \latin{et~al.}(2026)Kim, Xiong, Mahmood, Ramprasad, and
  Tran]{kim2026ai}
Kim,~C.; Xiong,~W.; Mahmood,~A.; Ramprasad,~R.; Tran,~H. AI-driven design of
  poly (ethylene terephthalate) replacement copolymers. \emph{Phys. Rev.
  Mater.} \textbf{2026}, \emph{10}, 033806\relax
\mciteBstWouldAddEndPuncttrue
\mciteSetBstMidEndSepPunct{\mcitedefaultmidpunct}
{\mcitedefaultendpunct}{\mcitedefaultseppunct}\relax
\EndOfBibitem
\bibitem[{ASKPOLY}()]{askpoly}
{ASKPOLY} at 
  \href{https://www.askpoly.io/}{https://www.askpoly.io/}\relax
\mciteBstWouldAddEndPuncttrue
\mciteSetBstMidEndSepPunct{\mcitedefaultmidpunct}
{\mcitedefaultendpunct}{\mcitedefaultseppunct}\relax
\EndOfBibitem
\end{mcitethebibliography}



\providecommand{\latin}[1]{#1}
\makeatletter
\providecommand{\doi}
  {\begingroup\let\do\@makeother\dospecials
  \catcode`\{=1 \catcode`\}=2 \doi@aux}
\providecommand{\doi@aux}[1]{\endgroup\texttt{#1}}
\makeatother
\providecommand*\mcitethebibliography{\thebibliography}
\csname @ifundefined\endcsname{endmcitethebibliography}
  {\let\endmcitethebibliography\endthebibliography}{}

\end{document}